\documentclass[iop]{emulateapj}

\shorttitle{quiescent galaxies in the high-redshift clusters}
\shortauthors{Lee et al.}

\begin{document}

\title{Evolution of Star-formation Properties of High-redshift Cluster Galaxies since $z = 2$}

\author{Seong-Kook Lee\altaffilmark{1,5}, Myungshin Im\altaffilmark{1,5}, 
Jae-Woo Kim\altaffilmark{1}, Jennifer Lotz\altaffilmark{2}, 
Conor McPartland\altaffilmark{3}, Michael Peth\altaffilmark{4}, Anton Koekemoer\altaffilmark{2}}

\altaffiltext{1}{Center for the Exploration of the Origin of the Universe, 
Department of Physics and Astronomy, Seoul National University, 
Seoul, Korea}
\altaffiltext{2}{Space Telescope Science Institute, 3700 San Martin Drive, Baltimore, MD 21218, USA}
\altaffiltext{3}{Institute for Astronomy, University of Hawaii, 2680 Woodlawn Drive, Honolulu, HI 96822, USA}
\altaffiltext{4}{Department of Physics and Astronomy, Johns Hopkins University, 3400 North Charles Street, Baltimore, MD 21218-2686, USA}
\altaffiltext{5}{sklee@astro.snu.ac.kr, mim@astro.snu.ac.kr}

\begin{abstract}

Using a stellar mass limited sample of $\sim 46,600$ galaxies 
($M_* > 10^{9.1}\,M_{\odot}$) at $0.5 < z < 2$, we show that the stellar mass, 
rather than the environment, is the main parameter controlling quenching of star 
formation in galaxies with $M_* > 10^{10}\,M_{\odot}$ out to $z=2$.
On the other hand, the environmental quenching becomes efficient at $z < 1$ 
regardless of galaxy mass, and it serves as a main star formation quenching 
mechanism for lower mass galaxies. 
Our result is based on deep optical and near-infrared 
imaging data over 2800 arcmin$^2$, enabling us to negate cosmic variance 
and identify 46 galaxy cluster candidates with $M \sim 10^{14}\,M_{\odot}$. 
From $M_* \sim 10^{9.5}$ 
to $10^{10.5}\,M_{\odot}$, the fraction of quiescent galaxies increases by 
a factor of $\sim 10$ over the entire redshift range, but the difference 
between cluster and field environment is negligible. Rapid evolution in the 
quiescent fraction is seen from $z=2$ to $z=1.3$ for massive galaxies 
suggesting a build-up of massive quiescent galaxies at $z > 1.3$. For galaxies with 
$M_* < 10^{10}\,M_{\odot}$ at $z < 1.0$, the quiescent fraction is found to 
be as much as a factor of 2 larger in clusters than in field, showing the 
importance of environmental quenching in low mass galaxies at low redshift.
Most high mass galaxies are already quenched at $z > 1$, therefore 
environmental quenching does not play a significant role for them, although the environmental quenching efficiency is nearly identical between high and low 
mass galaxies. 

\end{abstract}


\section{Introduction}

Galaxies define basic separate, independent entities in the universe 
and are the building blocks of her. 
Therefore, the study of their evolution throughout the history of the 
universe is the very core in our understanding of the universe. 

The $\Lambda$ cold dark matter (CDM) cosmological 
models say that galaxies form through the gravitational collapse of 
baryonic matters --- mostly in the form of gas ---  inside dark matter 
(DM) halos.
As gas collects within DM halos, it begins to cool and form stars, 
which gives galaxies the very glittering look as we observe them.
An interesting question regarding star-forming (SF) activity of galaxies 
is how this activity evolves and what governs this activity. 
Unlike the collapse of DM, which is mainly governed by gravity, SF 
activity is governed or affected by various baryonic physical processes 
--- including cooling of gas as well as feedbacks from stars, 
supernovae (SNe), or active galacitc nuclei (AGN) --- thus making 
a detailed understanding of the SF activity complicated.

Regarding the evolution of SF activity of galaxies, we know that galaxies 
in the local universe have much lower 
levels of SF activity than the ones in the past.
First, the global SF history, revealed from various galaxy surveys 
\citep{hop06a,cuc12,beh13,bur13,mag13,mad14} shows significant 
drop from redshift, $z \sim 1$ 
to $z \sim 0$ --- with a possible broad peak around $z \sim 1-3$.
While this shows the behavior of the collective SF activity in the universe, 
we can still speculate from this that the SF activity in individual 
galaxies must also decrease significantly.
Second, star-formation rates (SFRs) of normal SF galaxies at $z \sim 0$ 
are significantly lower than SFRs of higher redshift ($z \gtrsim 1$) SF 
galaxies with similar mass \citep[e.g.,][]{dad07,sob14}. 
This suggests an average drop in SF activities of individual galaxies 
with time toward $z \sim 0$. 
Lastly, a large fraction of local galaxies --- especially massive 
ones --- are red, quiescent, and form 
a tight red sequence \citep[e.g.][]{str01,bla03,bal04}. 
Therefore, we need to understand the physical mechanism(s) that 
can lower and eventually stop the 
SF activity of galaxies to explain this change of average SF 
properties of galaxies from high redshift to the local universe. 

At low redshift, it has been well known that SFR, color and morphology 
of galaxies, show strong dependence on their mass 
\citep[e.g.][]{kau03,jim05,bal06}, in such a way that less massive 
galaxies show more SF activities than massive ones. 
This clear mass-dependence seems to persist at higher redshift. 
By analysing the red-sequence luminosity function of a large sample 
($\sim 500$) of galaxy clusters at $z < 0.95$, 
\citet{gil08} found a dearth of red galaxies at faint luminosity at 
higher redshift. 
Recently, using the UltraVISTA \citep{mcc12} data, \citet{ilb13} have 
shown the stellar-mass dependence in SF quenching at $z < 1$ in a 
sense that stellar mass density of massive ($> 10^{11.2} M_{\odot}$) 
quiescent galaxies remains unchanged while less massive galaxies keep 
being quenched --- i.e., massive ones already quenched before $z \sim 1$, 
earlier than less massive ones. 
The origin of this mass-dependent SF cessation or quenching is still 
unclear. 
It could be the effect of feedback from AGN \citep{hop06,som08}, which is 
believed to selectively act for more massive galaxies. 
Another compelling explanation is the heating of accreted cold gas in 
massive halos \citep{bir03}. 

On the other hand, the SF activity --- or related properties, such as 
color or morphology --- shows environmental dependence as well. 
Locally, galaxies show distinct SF properties and colors in different 
environment \citep{lew02,kau04,bla05}.
Naturally arising questions are how and when these environmental trends 
have been developed. 
By analysing galaxy colors out to $z \sim 1.5$ from VIMOS-VLT Deep Survey 
(VVDS), \citet{cuc06} found that the color--density relation progressively 
weakens with increasing redshift and possibly reverses at $1.2 < z < 1.5$. 
\citet{gru11}, based on the Palomar Observatory Wide-Field Infrared 
\citep[POWIR;][]{con08} survey data in the DEEP2 field, also argued 
that there is no or weak correlation between galaxies' local number 
density and the color or the blue galaxy fraction for $0.4 < z < 1$ 
galaxies. 
In the case of the relation between SFR and density, \citet{coo08} found 
the reversal of the relation between SFR and overdensity at $z \sim 1$ 
using galaxy samples from the DEEP2 Galaxy Redshift Survey (GRS).
With GOODS data, \citet{elb07} also argued that the relation between SFR 
and density was reversed at $z \sim 1$, showing a sharp 
contrast to the local trend. 

On the contrary, \citet{coo06} argued that the mean galaxy environment 
shows a strong dependence on galaxy color at $z \sim 1$, similarly with 
what is locally found. 
\citet{pat09a}, analysing $z \sim 0.8$ galaxies, 
found that the total SFR--local-density relation still persists at this redshift. 
This environmental dependence of galaxy properties has been shown to 
persist even at higher redshift ($z < 2$) by \citet{qua12}, through the 
analysis of quiescent galaxy fraction of UKIDSS/UDS galaxies, even though 
the environmental trend of quiescent fraction becomes weaker with 
redshift. 
Combining data from several surveys --- UltraVISTA, 3D-HST \citep{bra12}, 
Cosmic Assembly Near-infrared Deep Extragalactic Legacy Survey 
\citep[CANDELS;][]{gro11,koe11}, and SDSS, \citet{tal14} 
have studied the evolution of quiescent galaxy fraction in groups 
over a wide range of redshift up to $z < 2.5$. 
Their finding is that quiescent fraction of group satellite galaxies 
is similar with background (i.e. field) galaxies at $z \sim 2$, while 
it increases faster than field galaxies at lower redshift down to $z \sim 0$. 
So, there seems to exist a certain degree of disagreement on how 
and how much the environmental trend changes as we go up to high redshift, 
even though there is a consensus on the fact 
that the environmental dependence weakens as redshift increases 
\citep[e.g.,][]{sco13}. 
Therefore, we have not yet reached to a firm conclusion about the rate 
or amount of evolution in the effects of environment on the formation 
and evolution of the quiescent galaxy population. 

In studying the role of environment in formation and evolution of 
quiescent galaxies, galaxy cluster offers a very unique laboratory. 
They define the densest galaxy environment, thus are expected to follow 
the highest density peaks in the universe. 
Besides the fact that denser structures collapse earlier in the $\Lambda$CDM 
paradigm, galaxies in these dense environments would also be 
subject to several physical processes --- such as ram-pressure stripping 
\citep{gun72}, strangulation \citep{lar80,bal00}, and harassment 
\citep{moo98} --- acting exclusively in group- or cluster-like environments.
In this regard, studying the stellar population properties of galaxies in 
galaxy clusters at various redshifts is a useful and crucial test bed 
in the investigation of the environmental effects on the formation and 
evolution of quiescent galaxy population. 

While there is a sharp contrast in galaxy properties between galaxies in 
clusters and in field locally, as the cluster environment is 
dominated by the red, quiescent galaxies \citep[e.g.,][]{dre80}, 
this picture seems to change as we go out to high redshift 
\citep{but84,dre97,pog01}.
Combining their field sample with the cluster samples from \citet{hol07}, 
\citet{van07} showed that the difference in the red galaxy fraction between 
the cluster- and the field environments increases with decreasing 
redshift --- i.e., the red galaxy fraction increases 
toward low redshift in the cluster environment while it remains nearly 
constant in the field environment. 
At higher redshift, $z > 1.5$, the story seems to become more complicated: 
Studies suggest that the galaxy SF properties (the SFR or the color) in the 
cluster environment at $z > 1.5$ can be different from those of cluster 
galaxies at lower redshift and that there can be cluster-to-cluster 
variation in the properties of these high-redshift proto-clusters.
\citet{gob11} analysed colors of red galaxies in 
a proto-cluster at $z \sim 2.1$ originally identified by their red 
$Spitzer$/IRAC colors, and find 
that these red galaxies have old stellar populations.   
On the other hand, \citet{zir08} found a well developed red-sequence in 
a $z \sim 2.2$ proto-cluster around a radio galaxy and suggested that 
many of these red galaxies are porbably dusty star-forming galaxies.
\citet{fas11} also reported the existence of the actively star-forming galaxies 
in the X-ray selected proto-cluster at $z \sim 1.6$. 
\citet{tra10} studied the color and SF properties of the member galaxies 
of $Spitzer$-selected $z \sim 1.6$ cluster \citep{pap10}.
They also found that the galaxies in this high redshift cluster, unlike the 
ones in lower redshift, are more dominated by blue galaxies. 
This, combined with the higher star-forming galaxy fraction within this 
cluster than in the lower redshift ($z \sim 0.3$), indicates that many 
cluster galaxies are still actively forming stars at $z \sim 1.6$.
The results of \citet{hil10}, based on the X-ray and mid-IR (MIR) 
observation, also show the strong SF activity of the cluster member 
galaxies at $z \sim 1.5$.    
\citet{str13} have found that there 
are massive star-forming galaxies along with passive ones even in the core 
region of the IRAC-selected $z \sim 2.1$ proto-cluster 
studied by \citep{gob11}. 
All together, these observational results point to the redshift range of 
$z > 1.5$ as the active formation era of the galaxy clusters as well as 
the epoch when many of cluster galaxies are still forming stars actively. 

Summarizing, both the mass and environment of galaxies seem to 
affect the quenching of SF activity of galaxies and the formation of 
quiescent galaxies. 
Also, the environmental dependence of SF activity of galaxies seems change 
with redshift. 
Therefore, more investigation about the timing (i.e., the redshift dependence) 
as well as the relative contribution from these distinct mechanisms 
--- mass versus environment --- is still needed.
\citet{pen10} have provided interesting results in this regard. 
Using SDSS \citep{yor00} and zCOSMOS \citep{lil07} data, they analysed 
the effects of mass as well as environment on the evolution 
and star-formation activity of galaxies at redshifts up to $z \sim 1$. 
They found that ``mass quenching" is more dominant for high stellar-mass 
($\gtrsim 10^{10}$ $M_{\odot}$) galaxies, while lower mass galaxies are 
more affected by ``environmental quenching". 
They also suggested that the combination of these two quenching mechanism 
can explain the evolution of stellar mass function of star-forming as 
well as quiescent galaxies. 
Analysing $z \sim 1$ H$\alpha$ emitters from the High-$z$ Emission Line 
Survey (HiZELS), \citet{sob11} have shown that the median SFR depends on 
galaxies' environment for galaxies with their stellar mass lower than 
$10^{10.6}\,M_{\odot}$, while there is no environmental dependence 
for more massive galaxies.

In this work, we extend this kind of investigation to higher redshift 
(up to $z \sim 2$), examining the effects of these two suggested 
drivers --- i.e. mass and environment --- in the formation and the 
evolution of quiescent galaxy population. 
It has been known that a tight red sequence already exists at $z \sim 1$ 
\citep[e.g.][]{imm02,bel04}. 
Therefore, investigation at higher ($z > 1$) redshift is crucial in 
catching the formation of these quiescent galaxies to place a meaningful 
constraints on their evolution.
Our work is based on the deep near-infrared (NIR) data from 
the United Kingdom Infrared Telescope (UKIRT) Infrared Deep Sky Survey 
(UKIDSS) Ultra Deep Survey (UDS) as well as deep optical data from the 
Subaru/XMM-Newton Deep Survey (SXDS), combined with mid-IR (MIR) data 
from the $Spitzer$ Space Telescope. 
Deep UDS NIR data are crucial in robust estimation of SF or stellar 
population properties of galaxies, breaking the degeneracy between old 
and dusty star-forming populations, thus in a study of quiescent galaxy 
population evolution at redshift as high as $z \sim 2$.
Our analysis is based on (1) an un-biased cluster detection (unlike, for 
example, the red-sequence method), (2) a relatively long redshift-baseline 
($0.5 \lesssim z \lesssim 2.0$) and a large spatial area ($\sim 665$ 
Mpc$^2$ at $z \sim 1$), and (3) robust estimation of SFR and stellar mass from SED-fitting. 
As demonstrated below, deep NIR data in the UDS field enables us the 
robust estimation of the stellar population properties of the galaxies 
up to $z \sim 2$.

We present the data set used in this work as well as the sample selection 
procedure in Section 2. In Section 3, we explain how the photometric redshifts as 
well as the stellar population properties of the UDS galaxies are estimated. 
We explain the identifying procedure of the high-redshift clusters and 
provide the properties of the selected cluster candidates in Section 4. 
We analyse the stellar population properties and the evolutionary trends 
of the quiescent galaxies, focusing on the drivers of this population 
in Section 5, and we summarize our results in Section 6.
We adopt the standard flat $\rm{\Lambda}$CDM cosmology, with 
($\Omega_{m}, \Omega_{\Lambda}$) = (0.3,0.7), and 
$H_{0}$ = 70 {\rm $km$ $s^{-1}$ $Mpc^{-1}$}. 
All magnitudes are given in AB magnitude system \citep{oke74} except when otherwise mentioned.

\section{Data and Sample}

The UDS (Almaini et al. in prep.), which is one of the five surveys in 
the UKIDSS \citep{law07}, provides a very deep NIR imaging dataset over an 
area of $\sim 0.77$ degree$^2$ located on the SXDS. 
The survey was carried out with the Wide Field Camera 
\citep[WFCAM;][]{cas07} on the UKIRT in three NIR broad-bands with the 
5-$\sigma$ limits of $J=24.3$, $H=23.3$, and $Ks=23.0$ (in Vega, DR10). 

The SXDS provides the deep optical data from the SUPRIMECAM on the Subaru 
telescope from $B$-band through $z'$-band. 
The 3-$\sigma$ depths of the Subaru data are $B=28.4$, $V=27.8$, $R=27.7$, 
$i'=27.7$, and $z'=26.7$. 
We use the released DR1 ($z'$-band detected) catalog \citep{fur08}, which 
is obtained from the SXDS DR1 release 
page\footnote{http://soaps.nao.ac.jp/SXDS/Public/DR1/}, 
for this work. 
The UDS region has also been observed by the $Spitzer$/IRAC as the SpUDS 
Spitzer Legacy Survey (PI:Dunlop), reaching $\sim 24$ magnitude 
(Channels 1 and 2).
We obtain the catalog at the Spitzer Legacy Survey 
archive\footnote{http://irsa.ipac.caltech.edu/data/SPITZER/SpUDS/}, and 
aperture-corrected total magnitudes are used for $Spitzer$/IRAC data. 
We also use the publicly available spectroscopic redshift data for about 
4000 objects (Simpson et al. in prep, Akiyama et al. in prep. and Smail 
et al. 2008)

We perform photometry on the UDS DR10 images ($J, H, Ks$) using 
SExtractor \citep{ber96} software. 
We use $9 \times 9$ convolution mask of a Gaussian PSF with FWHM of 5 pixels, 
and set the detection minimum area as 6 pixels. 
The detection and photometry are done in each band separately, and then 
the match between different bands is done with a matching radius of $1 \arcsec$ 
and with the $Ks$-band image as the reference image. 
About $0.8 \%$ objects (both in $J$- and $H$-band catalogs) have multiple 
matches, and the nearest one is selected as the best match. 
To construct the panchromatic spectral energy distributions (SEDs) from 
the optical to the MIR in the observed frame, we match, first, the Subaru 
optical and UDS NIR catalogs, and then with the SpUDS IRAC catalog, with a 
matching radius of $1 \arcsec$ and the position in $Ks$-band catalog 
is used as the reference position. 
For optical and NIR data, we use the auto magnitudes, and the aperture-corrected 
total magnitudes are used for IRAC data.  

We check the validity of using auto magnitude ($Mag_{{\rm auto}}$) 
in crowded region, such as clusters, by comparing auto magnitude values 
with aperture (circular diameter = $1\farcs5$) magnitude ($Mag_{{\rm aper}}$) 
in field and cluster regions. If $Mag_{{\rm auto}}$ values are biased in cluster 
regions, then we expect that $Mag_{{\rm auto}}$ to deviate from 
$Mag_{{\rm aper}}$ differently in cluster and field.  
At several magnitude bins ($Ks = 19.9, 20.9, 21.9, 22.9$), we compare the 
distributions of $Mag_{{\rm aper}} - Mag_{{\rm auto}}$ between cluster 
and field region through the Kolmogorov-Smirnov test (K-S test). 
At each magnitude bin, the maximum difference, $D$, was 0.073, 0.086, 0.071, 0.058, 
and the probability of null hypothesis, p, was 0.93, 0.62, 0.64, 0.94, showing 
that there is no systematic bias in auto-magnitude measurements that is 
specific in cluster region.

In our analysis, we use objects which are detected in both of optical (Subaru) 
and NIR (UKIRT) catalogs as our sample, but regardless of IRAC detection.
There are about $8.4 \%$ of $Ks$-band detected objects not detected in 
$z'$ band. The fraction is lower as $5.9 \%$ if we only consider the sources 
with $Ks \leq 24.9$. 
These $z'$-band undetected sources have, on average, redder IR colors than 
the detected ones.
The mean values of $(J-Ks)$ and $(J-3.6\mu m)$ are 1.0 and 1.8, respectively, while 
the corresponding values are 0.6 and 1.0 for $z'$-band detected objects.
And, about $25 \%$ of $Ks$- and $z'$-bands detected objects are matched with 
IRAC sources. 

From this `Subaru+UKIRT+Spitzer' multi-band photometric catalog, 
we first cull out stars either based on the spectroscopic classification 
or broad-band color ({\it BzK} or $J$-$K$). 
Specifically, from $\sim 200,000$ total objects, we exclude: (1) $\sim 550$ 
spectroscopically classified stars, (2) $\sim 22,000$ stellar candidates 
based on their location in the ($B$-$z'$)--($z'$-$Ks$) color--color plane, 
with the color criterion of $(z'-Ks) < 0.3 \times (B-z') - 0.4$, similarly 
with \citet{dad04} or \citet{han12}, and (3) further $\sim 500$ candidates 
with $J-Ks < 0$ and $Ks < 19.0$, based on the fact that galaxies and stars 
form clearly separate branches in this $J-Ks$ versus $Ks$ plane.

We need to cull out active galactic nuclei (AGN) from our sample because 
the measurements of photometric redshift and stellar population properties, 
which are crucial in our study, would be affected by the AGN activity. 
To remove AGN candidates from this star-removed, 
multi-band catalog, we use three steps: First, we exclude 77 objects 
which are classified as quasars in the spectroscopic catalogs. 
Next, we match our multiband catalog with the XMM-Newton point source catalog 
\citep{ued08} to remove any matched object.
At this step, about 220 objects are matched and excluded. 
Finally, for the objects with the matched IRAC photometry, we use the 
NIR $\&$ MIR color criteria suggested by \citet{mes12} to find out any 
remaining AGN candidates. 
The applied criteria are $(K - m_{4.5}) > 0$ and $(m_{4.5} - m_{8.0}) > 0$, 
where, $m_{4.5}$ and $m_{8.0}$ are the magnitudes in IRAC channel 2 
($4.5 \mu m$) and channel 4 ($8.0 \mu m$), respectively.
About 7,000 objects satisfy these color criteria, thus are excluded from 
the sample.
These color criteria, similarly with other IRAC color criteria, cannot 
distinguish between AGN and high-redshift ($z > 2$) galaxies. 
Therefore, by applying these criteria, we might lose some non-active 
galaxies at $z > 2$ as well. 
However, this does not affect our results, because our target redshift 
range does not exceed $z \sim 2$.

After trimming out AGN candidates as well as stars as explained, 
we apply the $Ks$-band magnitude cut ($Ks \leq 24.9$, 5 $\sigma$ detection limit) 
and the number of remaining galaxies is about 115,000 among which $\sim$1,400 
objects have spectroscopic redshift information. 
For these $\sim$115,000 objects, we derive photometric redshifts as well 
as stellar population properties through SED-fitting (Section 3).
After selecting galaxies within the redshift range of $0.45 \leq z \leq 2.1$ 
and applying the stellar mass cut of log $M_{*}/M_{\odot} \geq 9.1$ --- 
which is the $75 \%$ stellar mass limit at the highest redshift bin ($z \sim 2$) 
of our study --- our final sample size is 46,641. 

\begin{figure}[h]
\plotone{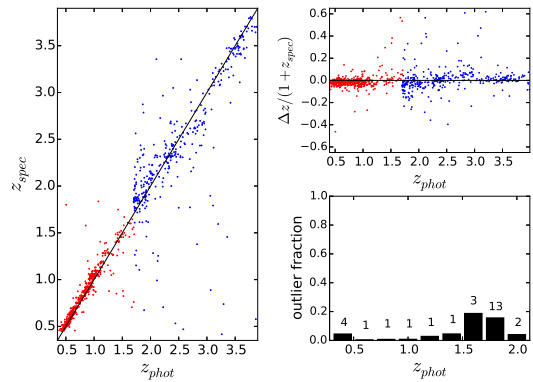}
\caption{{\bf (left)} Comparison of spectroscopic and photometric redshifts for 
the subset of UDS (red) and GOODS-S (blue) galaxies with available 
spectroscopic redshift. 
{\bf (upper right}) Photometric redshift discrepancy as a function of 
redshift. {\bf (lower right)} Outlier ($|\Delta z| / (1+z_{spec}) > 0.15$) 
fraction as a function of redshift. Numbers above each bar show 
the number of outliers at each redshift bin. \label{spzphz}}
\end{figure}

\section{Estimation of Photometric Redshifts and Stellar Population Properties}
\subsection{Photometric Redshift}

We derive photometric redshifts using the EAZY code \citep{bra08}. 
We take the $K$-band magnitude prior, which slightly improves the photometric 
redshift accuracy. 
We fit ten broad-band photometric data points from the observed frame optical 
(Subaru) to MIR ($Spitzer$/IRAC). 
During the fitting procedure, we require that at least five colors to be 
measured --- i.e. it is detected at least at six bands. 
Therefore, any galaxy which is not detected at more than five bands is not 
included in the following analysis. 
Only $\sim 0.14 \%$ of objects are excluded for this reason, and these are mostly 
faint objects --- $K > 24$ for $\sim 72 \%$ among these, and $z' > 26$ for $\sim 52 \%$. 

For a subset of galaxies for which spectroscopic redshifts are available and $Ks < 24.9$, 
we compare the spectroscopic and the photometric redshifts to estimate the 
accuracy of the photometric redshift (Figure~\ref{spzphz}, left panel). 
In UDS, there are not enough galaxies with $z_{spec} > 1.5$ for a fair 
estimation of photometric accuracy. 
To test the photometric redshift accuracy at $z \gtrsim 1.5$, we use the 
GOODS-S data, which carries similar coverage of broad-bands data 
with the UDS while having more abundant spectroscopic redshift samples at 
$z \gtrsim 1.5$. These GOODS-S spectroscopic data are from various 
spectroscopic surveys \citep{lef04,szo04,mig05,cim08,van08,pop09,bal10,sil10}.
We measure signal-to-noise of UDS photometric data as a function of magnitude 
at each band, and add additional noises (randomly following a Gaussian 
distribution) to the GOODS photometry to simulate the depth of the UDS data. 
Then, photometric redshifts are derived using these noise-added photometry.
Within the redshift range of $0.4 \leq z \leq 2.1$, 
the redshift range of interest of this work, there are $3 \%$ of objects 
whose redshift error, defined as $|\Delta z| / (1 + z_{spec})$, is greater 
than 0.15.
Removing these objects with catastrophic redshift errors, the mean 
error is 0.028. 

The right panels of Figure~\ref{spzphz} show $|\Delta z| / (1 + z_{spec})$ as well 
as the outlier fraction as functions of photometric redshift. 
We can see that our photometric redshift estimation is reliable 
within the redshift range of our study, $0.5 \lesssim z \lesssim 2$, 
but its uncertainty is larger at $z > 1.5$.
The outlier fraction increases at high redshift ($z > 1.5$), 
but it is kept at $<0.2$.
The reason for the increase in the photometric redshift uncertainty at 
$z > 1.5$ is that the spectral break at $\sim 4000$ \AA~ --- one of the 
clearest redshift indicators --- moves out of the optical regime at $z > 1.5$. 
When that happens, it becomes more difficult to sample the exact location of 
the rest-frame 4000 \AA~ break in the observed frame due to the gaps 
between NIR filters and the reduced number of available filters 
above 4000 \AA~ break. 
Also, the fraction of star forming galaxies are higher at $z > 1.5$ in 
the spectroscopic sample since it is easier to spot emission lines than 
absorption lines for faint galaxy spectra. 
Weak 4000 \AA~ breaks and strong emission lines of star forming galaxies 
make it difficult to trace the continuum shape, leading to a 
reduced accuracy in photometric redshifts \citep[e.g.,][]{dam09,ilb09,qua12,har14,yan14,kan15}. 
Not surprisingly, photometric redshift outliers tend to be star-forming galaxies, 
and some of them can move into the red sequence due to the large 
errors in photometric redshift.

In Figure 2, we show the $z_{spec}$ distribution of galaxies with several 
$z_{phot}$ bins ($z \sim 0.6, 0.8, 1.0, 1.2, 1.4$ and 1.6) with 
$|\Delta z| \leq 0.1$. 
Figure 2 shows that systematic offsets between 
photometric and spectroscopic redshifts are small ($\lesssim 0.1$ at some 
redshift bins).
The effect of the small systematic shift in redshift is 
negligible in our analysis. 
For example, a systematic offset of $\Delta z \sim 0.1$ causes a shift 
in the stellar mass estimate by $< 20 \%$.

\begin{figure}[h]
\plotone{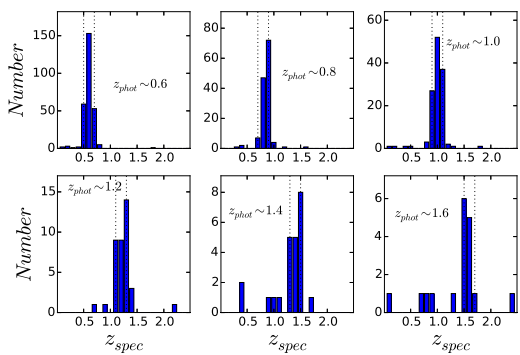}
\caption{Distributions of spectroscopic redshift of galaxies selected by their photometric 
redshift at various redshift bins. For example, at redshift 0.6, this figure (upper left 
panel) shows the distribution of spectroscopic redshifts of galaxies whose photometric 
redshifts are within $0.5 \leq z_{phot} \leq 0.7$. Systematic offsets are small, if there 
is any (< 0.1), between photometric and spectroscopic redshifts. The fraction of redshift 
outlier increases at higher redshift while the number of ourliers remain similar. \label{phzsprd}}
\end{figure}

\subsection{Stellar Population Properties from SED-fitting}\label{sedfit}

We perform the stellar population analysis through spectral energy 
distribution (SED) fitting methods for our sample, with the same procedure 
as explained in detail in \citet{lee09,lee10}. 
For redshift, we adopt $z_{spec}$ if available, but $z_{phot}$ otherwise. 

The stellar population synthesis models of \citet[][BC03, hereafter]{bru03} 
with Padova 1994 stellar evolutionary tracks are used for our SED-fitting 
analysis. 
We assume the \citet{cha03} initial mass function (IMF) with the lower and upper 
mass cuts of 0.1 and 100 M$_{\odot}$. 

As for the star-formation histories (SFHs), we use the delayed SFHs \citep{lee10}.
This form of SFH is shown to provide better estimation of stellar population 
properties than other forms, for example, exponentially declining ones 
\citep[e.g.][]{mar10,pap11,pfo12,lee14}.
Studying $z \sim 1$ galaxies from DEEP2 Redshift Survey \citep{new13}, 
\citet{pac13} have also shown that these galaxies have SFHs similar with 
this form. 
 
It has the following functional form. 

\begin{equation}\label{delaysfr}
\Psi (t,\tau) \varpropto \frac{t}{\tau^{2}} e^{-t / \tau},
\end{equation}
where, $\Psi (t,\tau)$ is the instantaneous SFR for a given set of 
$\tau$ and $t$.
The parameter $t$ is the time since the onset of the star-formation and $\tau$ is 
the time-scale parameter which governs how fast (or slowly) the SFR reaches its 
peak value before starting to decline. 
We allow the value of $\tau$ to vary within a very broad range from $0.1$ Gyr 
to 10.0 Gyr, and $t$ within 0.1 Gyr $\leq z \leq t_{H} (z)$, where $t_{H} (z)$ 
is the age of the Universe at redshift $z$. 

After extracting $\sim 4,000$ spectral templates from BC03 with these allowed values of 
parameters, we apply the \citet{cal00} attenuation law to model the dust attenuation 
of SEDs due to the inter-stellar dust. 
We vary the amount of attenuation, parametrized via $E(B-V)$ within 0.0 $\leq E(B-V) \leq$ 1.50 with the increment of 0.025. 
For the attenuation due to the line-of-sight neutral hydrogen in the inter-galactic 
medium (IGM), we apply the extinction law of \citet{mad95}.  
During the fitting, the redshift is fixed at the value of the spectroscopic redshift if available, otherwise at the photometric redshift. 

We apply the `grid-scanning' method which has been used in several previous works 
\citep{wik08,lee09,lee10,lee14}.
As explained in these references, our method of skimming the entire parameter space 
is less subject to the effects of any prior. 
From this SED-fitting results, we derive for each galaxy, the stellar mass ($M_*$), 
the SFR, the mean age of the stellar population, the amount of dust attenuation 
($E(B-V)$), as well as the star-formation history --- parametrized in terms of 
$\tau$ and $t$. 
The SFR used in this work is defined as the SFR averaged over recent 100 Myr, based 
on the reasoning outlined in \citet{lee09}.
We also derive the rest-frame $u-g$ color --- will be denoted as 
$(u-g)_{0}$ from now on --- for each galaxy from its best-fit BC03 template.  

\begin{figure}[h]
\plotone{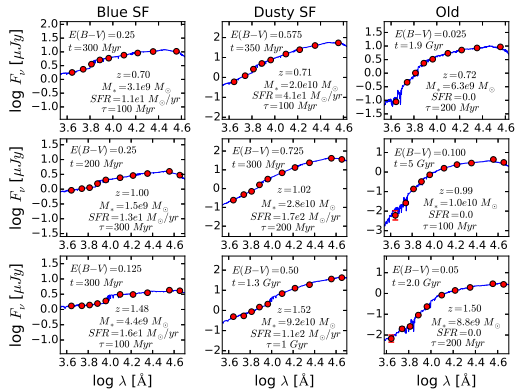}
\caption{Examples of observed SEDs and the best-fit BC03 spectrum for blue SF 
({\bf left column}), dusty SF ({\bf middle column}), and old passive 
({\bf right column}) galaxies at $z \sim 0.7$, 1, 1.5. 
In each panel, red circles show the observed fluxes from $BVRiz$ (Subaru) to 
$JHKs$ (UKIRT) to MIR (Spitzer) upto IRAC Channel 2 (4.5 $\mu m$). 
Solid blue curve is the best-fit BC03 spectrum found through SED-fitting. 
We can see that blue SF galaxies have bluer colors than red, dusty SF galaxies throughout 
the shown wavelength range as well as that dusty SF galaxies and old quiescent ones can 
be discriminated reliably thanks to the deep and wide wavelength coverage of the data.
Wavelength ($x$-axis) is in logarithmic scale of \AA, and flux is in log-scale of $\mu Jy$. \label{sedexm}}
\end{figure}

The broad wavelength coverage out to IR regime as well as the deep optical 
data from Subaru in the UDS field enables us to estimate the stellar 
population properties of high-redshift galaxies with high confidence 
and distinguish dusty SF galaxies from old galaxies among red galaxies. 
Figure~\ref{sedexm} shows the examples of the SED-fitting results for a representative 
set of galaxies at various redshifts. 
Also shown are the key SED-fit parameters. 
In this figure, we show the observed galaxy SEDs (red circles) as well as 
the best-fit BC03 spectra for typical blue SF galaxies 
(panels in left column), dusty SF ones (panels in middle column) and old 
quiescent ones (panels in right column) at redshifts $z \sim 0.7$, 1, 1.5.

Here, we define ``quiescent galaxies" as galaxies with 
sSFR $< 1/[3 t(z)$] yr$^{-1}$, where $t(z)$ is the 
age of the universe at $z$.
This cut was used in several previous works \citep[e.g.,][]{dam09}, 
and approaches the local value for sSFR cut at $z \sim 0$ 
\citep[e.g.,][]{gal09,ko14}.
Under a delayed SFH model (i.e., Equation (1)), 1/[$3 t(z)$] corresponds to 
$\sim 1\%$ of the peak SFR.
While this sSFR cut select galaxies with low SF activity in comparison to the 
average SF galaxies at that redshift, the selected galaxies can have 
SFRs that are not quite ``quiescent". 
For example, at $z \sim 2$, $M_{*} \sim 10^{11} M_{\odot}$ galaxies 
at the sSFR cut have $\sim 10 M_{\odot}$/yr --- i.e., SFR of starburst 
galaxies in the local universe.
If we fit the SEDs of a 1 Gyr-old, passively evolving galaxies after a 
spontaneous burst (a truly quiescent galaxy at high redshift),  
our procedure returns sSFR $\sim 10^{-11}$ yr$^{-1}$.
We check the MIPS 24 $\mu$m detection of our galaxies using SpUDS MIPS 
catalog. 
We find that small fraction ($\sim 5 \%$) of red SF galaxies is 
detected, while no quiescent galaxy is detected at 24 $\mu$m. 

\begin{figure}[h]
\plotone{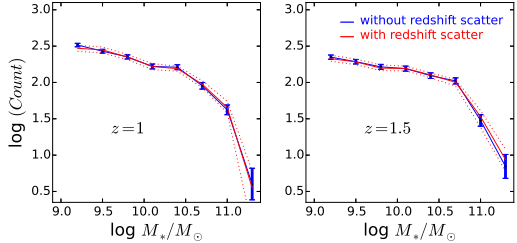}
\caption{Galaxy number counts (and error) at stellar mass bins (blue) and the 
same ones but with photometric redshift scatter effect included (red) at 
$z = 1$ (left) and $z=1.5$ (right). 
Red solid and dotted lines in each panel show the mean and envelope (i.e. minimum and maximum) 
of Monte Carlo simulation (100 times) which include the redshift scatter. 
We can see that the effect of photometric redshift error is not significant in the estimation of the stellar mass function. \label{lmszerr}}
\end{figure}

We test how the error in photometric redshift measurement may affect the 
estimation of stellar masses of galaxies using Monte Carlo simulation. 
Specifically, first, we scatter the redshift of each galaxy assuming 
a Gaussian distribution of redshift error. 
Then we repeat SED-fitting by fixing the redshift at this scattered value. 
We repeat this procedure 100 times each galaxy at $z \sim 1$ 
and $z \sim 1.5$, within a redshift range $\pm 0.02$. 
The results of this simulation are given in Figure~\ref{lmszerr}, where we 
show the number counts of galaxies at stellar mass bins. 
The resulting number distributions with scattered redshift (red) show 
no significant difference from the ones without scatter (blue) at 
both redshift bins. 
For individual galaxies, the mean and the standard deviation of 
$\Delta M_{*} / M_{*,true}$ is $0.02 \pm 0.11$ and $0.01 \pm 0.09$, at 
$z \sim 1$ and $\sim 1.5$, respectively.
Here, $\Delta M_{*}$ is $M_{*,MC} - M_{*,true}$, and $M_{*,true}$ is the 
measured stellar mass without redshift scatter.
From this, we can ensure that the stellar mass is not 
significantly affected by the error in photometric redshift measurement. 

\section{Massive Structures of Galaxies}

\subsection{Stellar Mass Limit}

Based on the stellar mass derived from the SED-fitting, we select galaxies with their 
stellar masses greater than $10^{9.1}~M_{\odot}$ (which is the $75 \%$ stellar mass 
limit at the highest redshift range, $z \sim 2.0$) within the redshift 
range of $0.45 \leq z \leq 2.1$. 

\begin{figure}[h]
\plotone{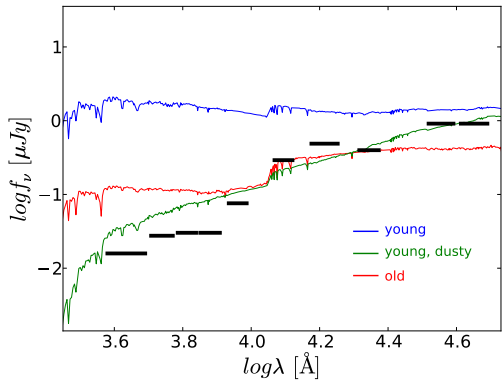}
\caption{Depths of the datasets used for our study, in comparison to different 
types of galaxies.
Black horizontal bars show the flux limit of the UDS in 5 Subaru filters 
($B$, $V$, $R$, $i'$, $z'$), 3 UKIRT filters ($J$, $H$, $Ks$), and 2 IRAC 
channels ($3.6 \mu m$, $4.6 \mu m$). 
All the galaxy templates are normalized to have stellar mass of $10^{9.1} M_{\odot}$. 
Solid blue curve shows the spectrum of a young galaxy template with $t = 200$ Myr 
with no dust attenuation. 
Solid green curve is for the same galaxy template but with high dust attenuation 
value of $E(B-V) = 0.4$. 
Red curve shows an old, quiescent galaxy template with $t = 2$ Gyr --- maximally 
possible age at $z=2$ if we assume the formation redshift as $z_{form}=5$. 
This figure shows that we can detect with our data set galaxies at $z \sim 2$ 
with stellar mass as low as $\sim 10^{9.1}\,M_{\odot}$.\label{sedlim}}
\end{figure}

We have checked if the magnitude limit is deep enough to detect galaxies with 
stellar mass of $10^{9.1} M_{\odot}$ out to $z=2.0$, using galaxy template 
spectra of various galaxy types extracted from the BC03 library. 
In Figure~\ref{sedlim}, we show these spectra --- young (solid blue curve), 
dusty young (solid green curve), and old (solid red curve) --- with stellar mass of 
$10^{9.1} M_{\odot}$ and redshifted to $z=2.0$. 
We also show the limiting flux of the UDS data set in 5 Subaru 
optical filters ($B$, $V$, $R$, $i'$, and $z'$), in 3 UKIRT filters 
($J$, $H$, and $Ks$) and in 2 IRAC channels ($3.6 \mu m$ and $4.5 \mu m$) 
shown as black horizontal bars. 
We find that SEDs of all the galaxy types are above the detection limits in multiple 
bands, demonstrating that our dataset is nearly suitable for selecting galaxies 
with log ($M_{*}/M_{\odot}$) $> 9.1$.  

\begin{figure}[h]
\plotone{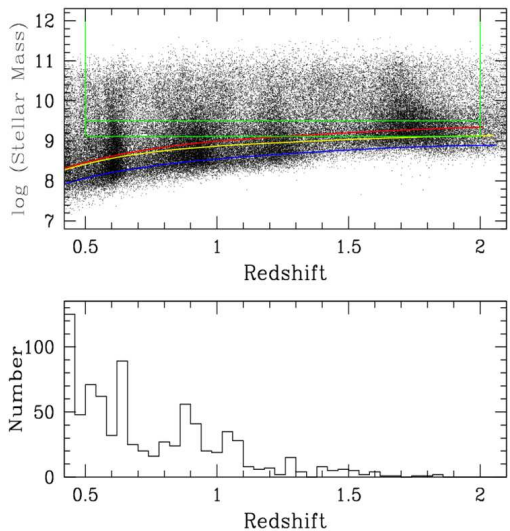}
\caption{{\bf (Top)} Redshift versus stellar masses derived from the SED-fitting. 
Green lines show the redshift range (vertical) as well as the stellar mass limits 
(upper horizontal: near $100\%$ at $z < 2$; lower horizontal: near $100\%$ at $z < 1.5$ 
and $75\%$ at $z < 2$) of our sample.
Solid curves shows the stellar mass at different redshift of quiescent 
(yellow and red) and SF (blue; $z_f = 3$) BC03 galaxy templates with $K = 24.9$. 
The yellow and red curves correspond to quiescent galaxies 
with reasonable ($z_f = 5$) and extreme ($z_f = 10$) formation epoch, respectively. 
This figure illustrates that we can detect reasonably old galaxies with log 
($M_*/M_{\odot}$) = 9.5 up to $z=2$, 
while we may miss some extremely old, low-mass galaxies, if any. 
{\bf (Bottom)} Distribution of spectroscopic redshift. It shows peak near 
$z \sim 0.6$ and $\sim 0.9$ similarly as total (photometric redshift) sample 
as shown in top panel. The number of spectroscopic sample quickly drops 
at $z > 1$. \label{zlms}}
\end{figure}

Figure~\ref{zlms} shows the redshift versus stellar mass of galaxies 
in the $K$-band limited sample. 
In this figure, the green vertical lines show the sample redshift range of 
this study and the green horizontal lines show the stellar mass limits.
Blue curve shows the stellar masses at different redshifts of star-forming galaxy 
templates from BC03 with formation redshift of $z_f = 3$. 
Yellow curve corresponds to quiescent galaxy templates with $z_f = 5$, which 
is a reasonable formation redshift for faint/low-mass galaxies, while red curve 
is for galaxy templates with extremely high formation redshift ($z_f = 10$).
As shown in this figure, within the given redshift range, we can detect 
log ($M_{*}/M_{\odot}$) $\gtrsim 9.1$ galaxies (corresponding lower green horizontal 
line) either quiescent (yellow curve) or star-forming (blue curve). 
It is still possible that we may miss some low mass 
($ < 10^{9.5} M_{\odot}$) galaxies with extremely old ($z_f \sim 10$) 
population, even though it is reasonable to assume that there are only few, if any, 
galaxies with these extreme properties in reality. 
For this reason, we apply the stellar mass cut of $10^{9.5} M_{\odot}$ (upper green 
horizontal line) when we deal with quantities which might be affected by 
this missing of low-mass old galaxies.
Within this redshift range and above the stellar mass limit 
(log ($M_{*}/M_{\odot}$) = 9.1), we have 46,641 galaxies. 

\subsection{Method and Procedure}

Our method of finding galaxy cluster candidates is based on the photometric 
redshift and similar with the method used in \citet{kan09,kan15}. 
Following is the detailed explanation of the procedure we use to 
identify the cluster candidates.
First, we divide the projected sky area into retangular grids with 
a width of $12 \arcsec$, which corresponds to $\sim 96$, 102, and 100 
kpc at $z = 1$, 1.5, and 2. 
Then, at a given redshift bin and at each grid point, we count the galaxies with 
log ($M_{*}/M_{\odot}) \geq 9.1$ within a radius of $r \leq 700$ kpc 
($\sim 1.5 \arcmin$ at $z \sim 1$) from the grid point within the redshift 
range of the typical redshift error in the UDS (i.e., 
$\Delta z = \pm 0.028 \times (1+z)$).
We repeat this procedure at redshift bins with an increment of $\Delta z$ = 0.02. 

Next, we find the mean ($\bar{N}$) and the standard deviation ($\sigma_N$) of the 
galaxy number counts ($N$) through the Gaussian fitting at each redshift bin. 
Then, we identify the $over$-$dense$ grid points with their galaxy number counts 
that exceed the 4-$\sigma$ level from the mean, i.e., 
\begin{equation}\label{neqn}
N ~\geq ~\bar{N} + 4 \times \sigma_N.
\end{equation} 

If we see the spatial distribution of these $over$-$dense$ grid points with extreme 
galaxy number counts ($\geq 4 \sigma$), we find that some points are connected to 
other points making large structure while some points are relatively isolated.
In our case, the spacing between the grid points ($\sim 100$ kpc) is much smaller than 
the typical size of galaxy clusters ($\sim 1$ Mpc). 
Therefore, a cluster of galaxies would appear as a structure with connected 
$over$-$dense$ grid points.  
For example, if 9 grid points are connected with each other with a square shape 
(i.e., $3 \times 3$) or if 10 grid points are connected in a rectangular shape 
(i.e., $2 \times 5$), this connected structure would represent an overdense structure 
with a radius of $\sim 850 - 900$ kpc.
Based on this reason, we exclude any $over$-$dense$ grid points when the 
number of connected grid points is smaller than 10 at each redshift bin. 
Then, among the remaining connected structures, we only select the structures 
as our final (proto-)cluster candidates only when these structures are identified at 
the same sky location (within the displacement, $d \leq 1.5$ Mpc) in at least 
three successive redshift bins. 
In other words, if the connected structure is found only in one or two  
redshift bin(s), we discard this structure.
This criterion has been applied to minimize any false detection of cluster candidate.
By applying these two, rather conservative criteria in selecting cluster candidates, 
we try to make our selection of cluster sample as robust as possible. 
By comparing with a mock catalog from the GALFORM simulation \citep{mer13}, 
we find that our cluster finding method is nearly complete at selecting 
$M_{h} > 10^{13.8} M_{\odot}$ clusters at $z \sim 0.9$ and $z \sim 1.5$. 
A detailed description of this method applied on simulation data 
can be found in \citet{kan15}. 

\begin{figure}[h]
\plotone{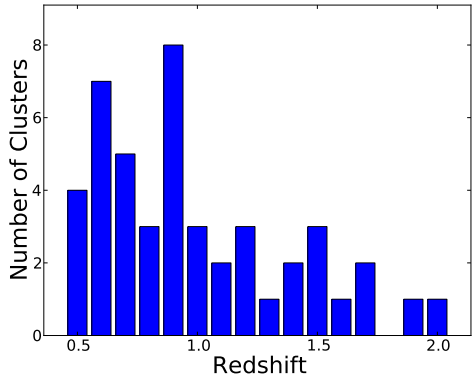}
\caption{The number of identified cluster candidates within the redshift range 
of $0.5 \lesssim z \lesssim 2.0$. \label{clnum}}
\end{figure}

\subsection{Galaxy Cluster Candidates}

Following the procedure as explained in the previous section, 
we find 46 cluster candidates within the redshift range of 
$0.5 \lesssim z \lesssim 2.0$. 
Among these, 19 clusters are the ones already identified by other authors 
\citep{vbr06,vbr07,fin10,pap10} in the same UDS-field, and 
27 clusters are newly identified in this work. 
Figure~\ref{clnum} shows the number of the cluster candidates 
at each redshift bin. 

In Table 1, we present the basic properties of these cluster candidates, including 
the number of member galaxies (Column 4), total stellar mass of member galaxies 
(Column 5), and the overdensity measure, defined as $(N - \bar{N}) / \sigma_N$ 
(Column 6). 

\begin{figure}[h]
\plotone{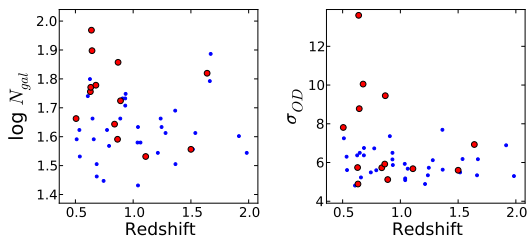}
\caption{Total number of member galaxies (left panel) and the maximum overdensity factor 
(right panel) of our galaxy cluster candidates. Red circles are 13 galaxy clusters 
identified by \citet{fin10}. \label{clsigng}}
\end{figure}

After finding the cluster candidates, we define the spatial center and 
the redshift of the candidate clusters through the following procedure. 
At each redshift bin, the mean position, weighted by the galaxy 
count ($N$), is found for each connected over-dense structure.
Next, the center and the redshift of each cluster candidate is assigned 
in a similar manner, i.e, as the $N$-weighted mean center and redshift, for 
each of the connected redshift bins. 
Then, we find the galaxies within the radius, $r \leq 1$ Mpc --- which corresponds 
to, or is slightly smaller than, a typical radius of a galaxy cluster --- from 
the center and within the redshift interval, $|\Delta z| \leq 0.028 \times (1+z)$ 
--- which is the typical redshift error of our data.
Finally, we re-calculate the center and the redshift of each cluster candidate 
as the stellar-mass weighted mean center and redshift of the member galaxies 
belonging to the overdense area. 
These center and redshift are the values listed in Columns (1), (2) and (3) 
in Table 1.
The number of member galaxies varies within the range between $\sim$30 and 
$\sim$90 (the left panel of Figure~\ref{clsigng}).
The right panel of the same figure shows the over-density measure, defined as 
($N - \bar{N}$)/$\sigma_{N}$, and shown as $\sigma_{OD}$ ($y$-axis) in the figure.
Most clusters have $\sigma_{OD}$ values between 4 and 8, and 4 clusters at 
redshift $z < 1$ have higher $\sigma_{OD}$. 

\begin{figure}[h]
\plotone{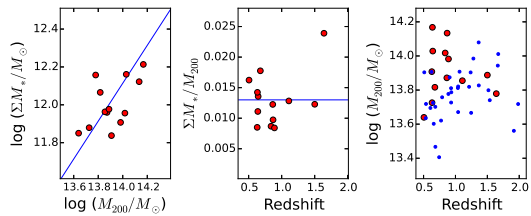}
\caption{{\bf Left:} Correlation between the cluster halo mass ($M_{200}$) 
and the total stellar mass ($\Sigma M_{*}$) of 13 galaxy 
clusters which are also identified by \citet{fin10}. 
Cluster halo masses are from \citet{fin10}. As can bee seen here, there exists 
positive correlation between the total stellar mass and the halo mass. 
Blue line shows the constant $\Sigma M_{*}/M_{200}$ ratio of 0.013. 
{\bf Middle:} The stellar-to-halo mass ratios and their redshifts for the same 
13 galaxy clusters. 
{\bf Right:} Cluster halo masses and their redshifts for all 46 galaxy clusters. 
Red circles are the same 13 galaxy clusters shown in left two panels. 
Blue small circles are the remaining 33 clusters. For these 33 clusters, we 
assume the stellar-to-halo mass ratio of 0.013. \label{clmhms}}
\end{figure}

Out of 46 clusters, 13 clusters are also found by \citet{fin10} based on the $X$-ray observation, 
and they provide the total mass ($M_{200}$) measured from the $X$-ray data, and 
the corresponding $r_{200}$.
We compare the sum of the stellar masses of the cluster member galaxies and their total 
(halo) mass for these 13 clusters, and these two mass measures ---  halo mass ($M_{200}$) and total stellar mass ($\Sigma M_{*}$) --- show a positive correlation 
as shown in the left pane of Figure~\ref{clmhms}.
In the middle panel of this figure, we show the redshift versus 
the mass ratio ($\Sigma M_{*}/M_{200}$) of clusters. 
Here, we cannot see any clear redshift-dependent trend, and the mean mass 
ratio is 0.013, shown as a blue horizontal line in this panel.  
The right panel shows the redshift versus $M_{200}$ of our 46 clusters.
Here, red large circles are the 13 clusters with their $M_{200}$ taken from 
\citet{fin10}. 
Blue small circles are the remaining 33 clusters, for which we assign 
a halo mass assuming the stellar-to-halo mass ratio of 0.013.
These clusters have halo masses of $3 \times 10^{13}$ to 
$2 \times 10^{14}\,{\rm M_{\odot}}$, and total stellar masses of 
$4 \times 10^{11}$ to $2 \times 10^{12}\,{\rm M_{\odot}}$. 

\section{Color and Star Formation Properties}

In this section, we take a close look at the color and SF properties of 
galaxies belonging to the cluster candidates, compare these properties 
with those of field galaxies in the same UDS field, and investigate their 
evolutionary trend. 

\begin{figure}[h]
\plotone{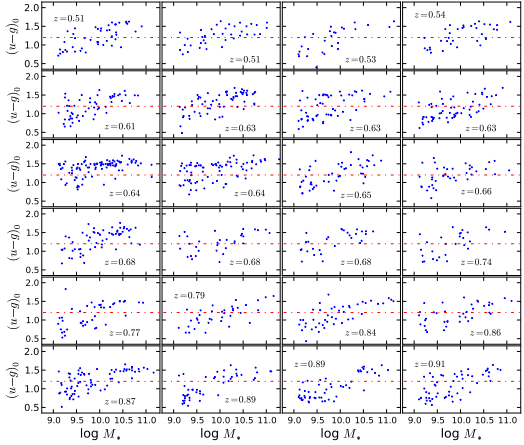}
\caption{The $(u-g)_{0}$ versus $M_*$ plots of cluster member galaxies. Each panel shows the 
color-mass diagram for each individual cluster candidates. In each panel, the redshift of the cluster 
is given. Red dot-dashed line in each panel shows the dividing line 
($(u-g)_{0}$ = 1.2) between the red and the blue galaxies. \label{colms1}}
\end{figure}

\begin{figure}[h]
\figurenum{10}
\plotone{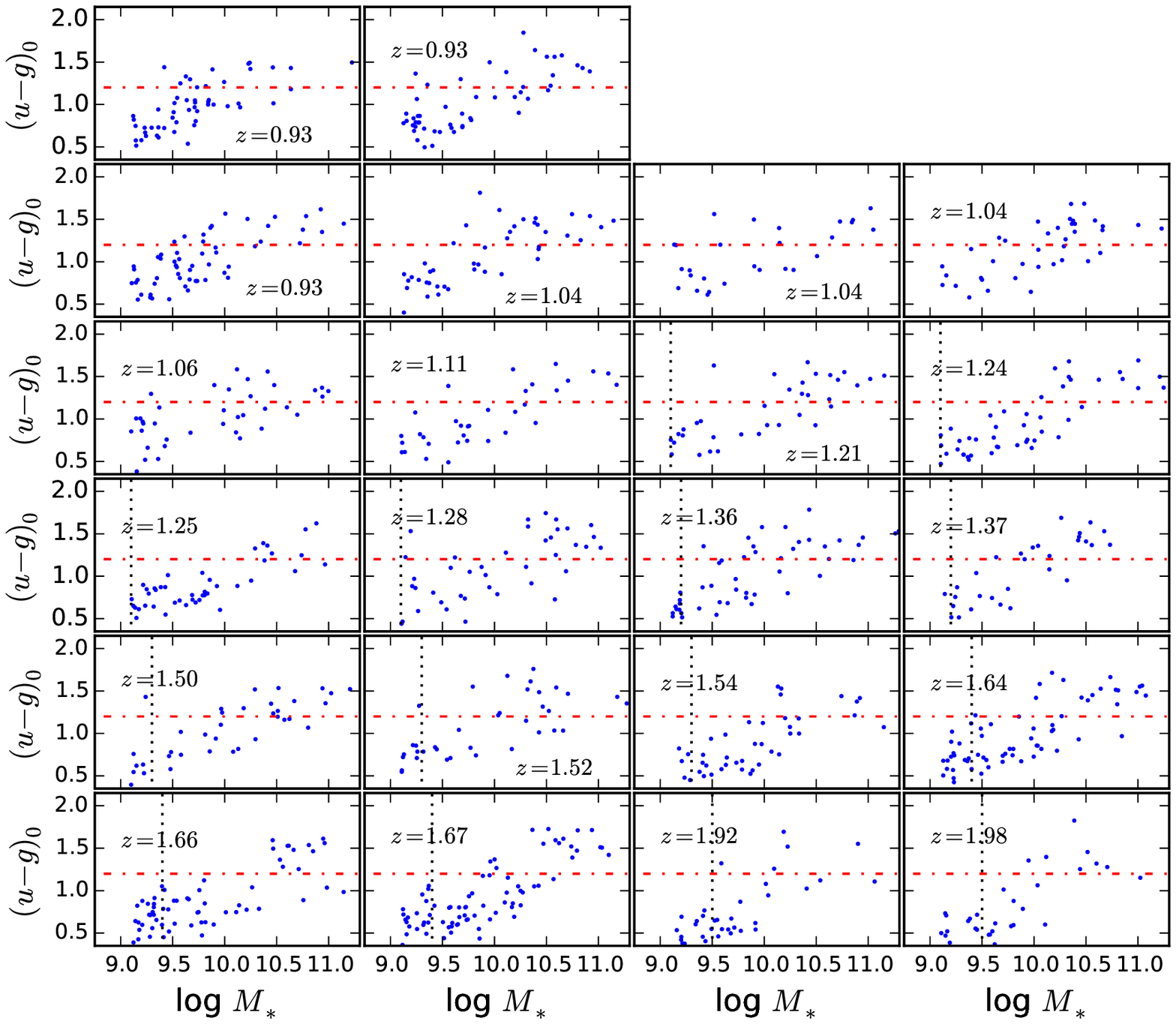}
\caption{Continued. Black vertical dotted line in each panel shows the stellar 
mass limit (shown as red curves in Figure 5) at corresponding redshift. 
We do not show this limit when it is smaller than 9.1. \label{colms2}}
\end{figure}

\subsection{Optical Color and its Evolution}\label{colnevol}

\subsubsection{Color of Cluster and Field Galaxies}

Figure~\ref{colms1} shows $(u-g)_{0}$ color versus stellar mass diagrams 
of member galaxies of each cluster candidate. 
As shown in this figure, the red sequence is formed throughout the whole 
redshift 
range, while the fraction of the galaxies with red color  
decreases with increasing redshift. 
Hereafter, we refer the galaxies with $(u-g)_{0} > 1.2$ as red galaxies, 
and the ones with $(u-g)_{0} < 1.2$ as blue galaxies.
In clusters, a significant $sequence$ of red galaxies starts to appear 
at redshift as high as $z \sim 1.6$.

At the very highest redshift ($z > 1.9$), only several massive galaxies 
have red ($> 1.2$) colors while progressively lower mass galaxies 
join the red sequence with decreasing redshift. 
At $z \lesssim 1$, we can see that the red sequence is well extended 
down to very low stellar mass (log $(M_{*}/M_{\odot}) \lesssim 9.5$) in 
most of the clusters.
This is a $cluster$ version of downsizing phenomenon, which is also shown 
in the study of two galaxy clusters (at $z \sim 0.8$ and $\sim 1.2$) 
by \citet{nan13}. 
We confirm this trend of mass-dependent timing of red sequence formation 
with a significantly larger sample. 
A similar, luminosity-dependent trend in a deficit of red galaxies was 
suggested earlier by \citet{gil08}. 
Please note that we can detect red galaxies down to a low-mass limit 
($= 10^{9.1} M_{\odot}$) throughout the redshift range ($0.5 < z < 2.0$), 
thanks to the very deep optical data from SXDS.
Therefore, the dearth of low-mass red galaxies at highest redshift bins 
is real --- i.e., not affected by the magnitude limit of the data. 

\begin{figure}[h]
\plotone{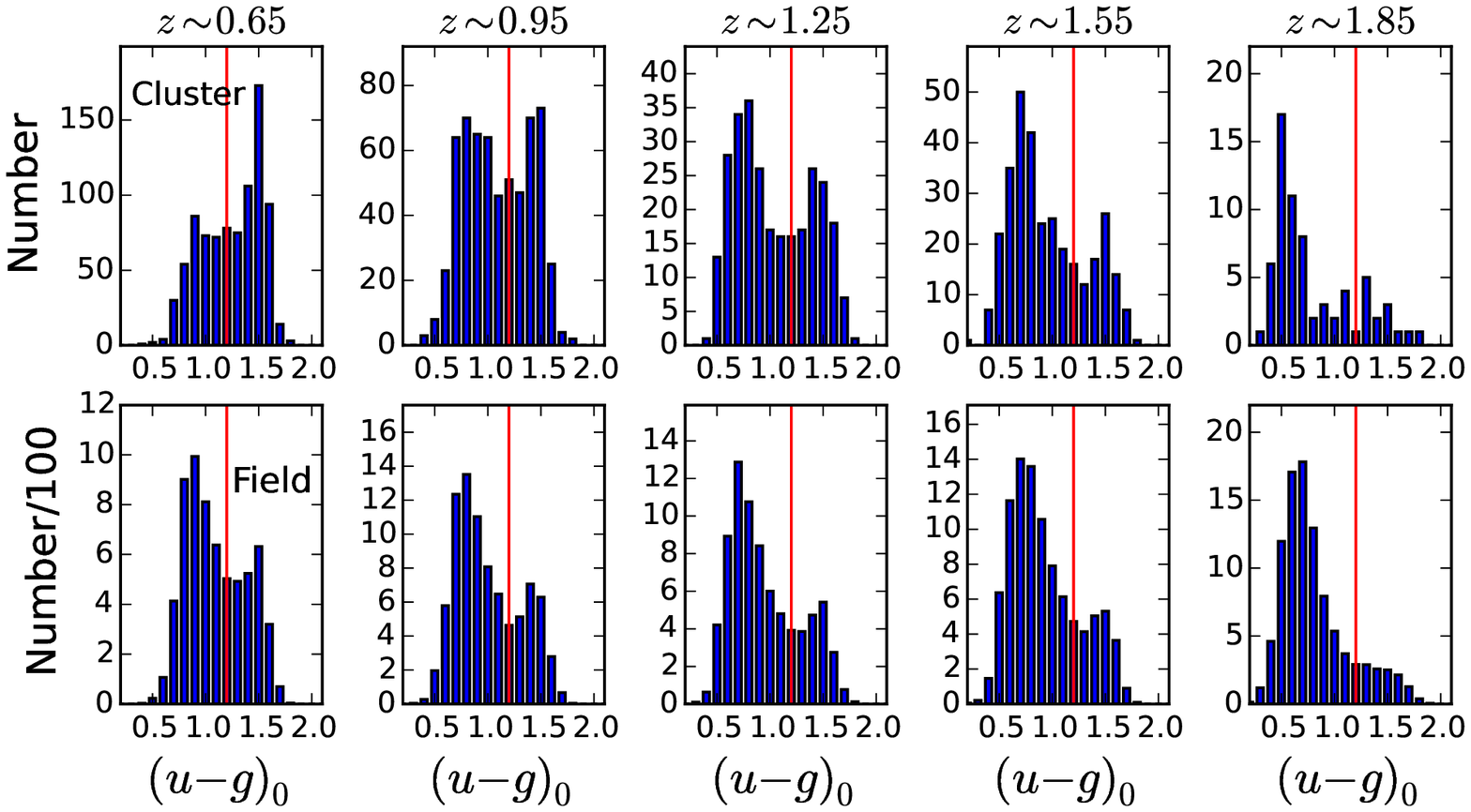}
\caption{{\bf Top:} The $(u-g)_{0}$ color distributions of cluster galaxies 
in five redshift bins. The cluster galaxies are summed in each redshift bin. 
{\bf Bottom:} The $(u-g)_{0}$-distributions of the field galaxies in the 
same redshift bins as in the top row. The rest-frame color is derived from the 
best-fit BC03 template for each galaxy. Red vertical line in each panel shows the 
dividing line ($(u-g)_{0}=1.2$) between red and blue galaxies. Both in cluster- and 
field-environment, color distributions show clear bimodality, except at the highest 
redshift bin ($z \sim 1.85$). 
There is no clear difference in color distribution between clusters and field at 
the highest redshift while the distributions are clearly distinguished at the lowest 
redshift ($z \sim 0.65$). 
The evolution of color distribution in clusters starts to deviate from the field ones 
from $z \sim 1.25$. \label{coldstclfl}}
\end{figure}

In Figure~\ref{coldstclfl}, we present the distributions of $(u-g)_{0}$ color 
at five redshift bins ($0.5 \leq z < 0.8$, $0.8 \leq z < 1.1$, 
$1.1 \leq z < 1.4$, $1.4 \leq z < 1.7$, and $1.7 \leq z < 2.0$) for cluster (top row) 
and field galaxies (bottom row). 
As can be seen in the color-$M_{*}$ diagrams, $(u-g)_{0}$ color 
shows a clear bimodal distribution, which suggests that the color transition 
of galaxies occurs in a short time scale. 
At redshifts greater than $z \sim 1.4$, the clusters are dominated by blue 
galaxies, while at $z \leq 0.8$, clusters are $red$-$dominated$ 
--- i.e., majority of the cluster galaxies are red at $z \leq 0.8$. 
But, red galaxies exist even at the highest redshift bin ($z \sim 1.85$). 
We can also observe the reddening of the peak color of blue galaxies with 
decreasing redshift, which reflects the evolution of the average stellar 
population of blue galaxies with time. 

When we compare the $(u-g)_{0}$ color distributions 
of the cluster galaxies (top row) and of the field galaxies (bottom row), 
we can see that the color distribution is similar 
in clusters and in field at the highest redshift bin ($z \geq 1.7$). 
At $z \leq 1.4$, the color distribution in clusters becomes 
distinguished from that of the field galaxies.
Clusters become to be more $red$-$dominated$ compared to field
at the same redshift.  
At the lowest redshift bin, blue galaxies are still the major population 
in field, while red galaxies clearly dominate in clusters.
From this, we can conclude that the color transition of galaxies are accelerated 
in clusters at redshift lower than $z \sim 1.4$.

\subsubsection{Color Evolution in Clusters and Field}

\begin{figure}[ht]
\plotone{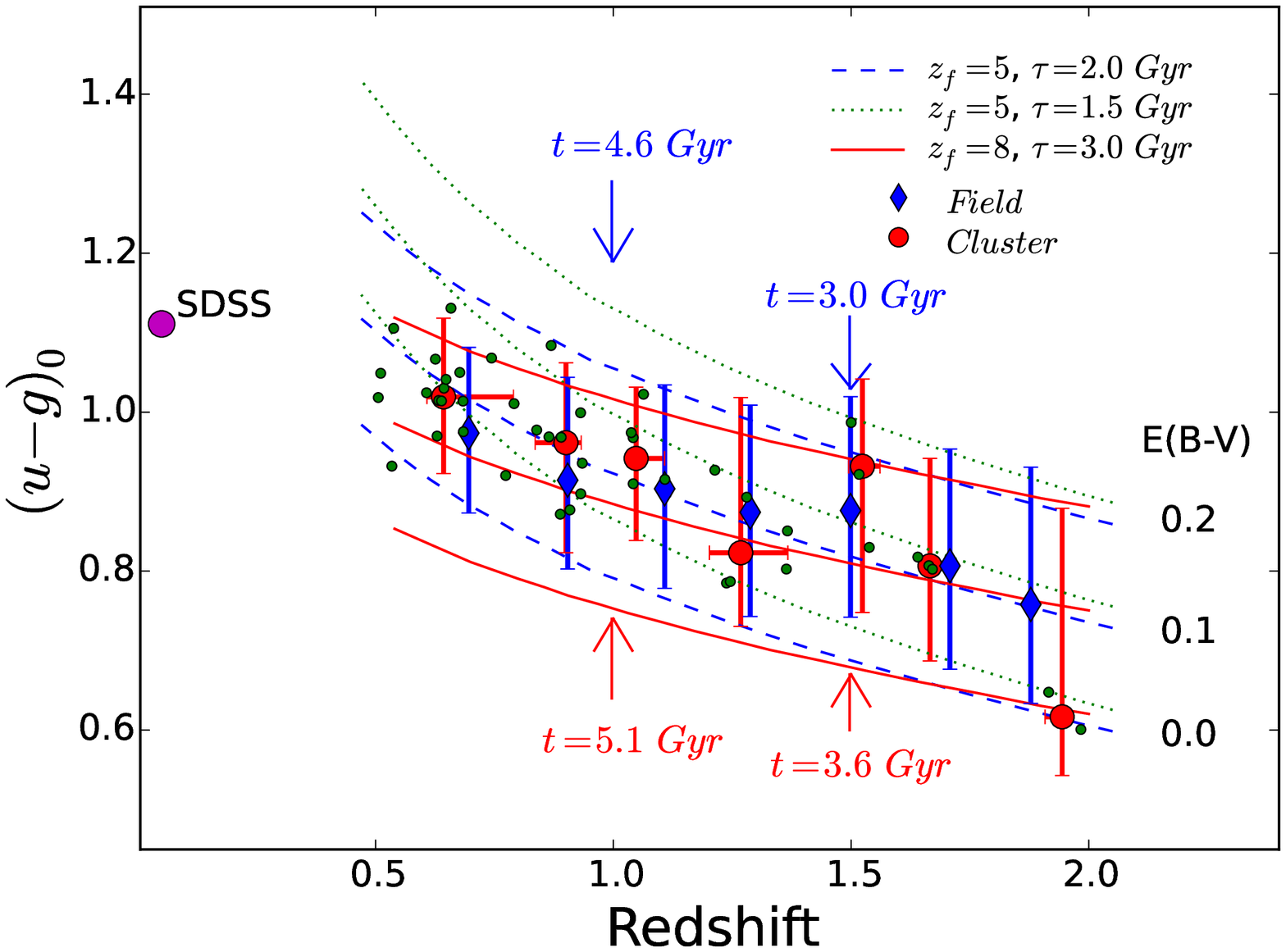}
\caption{The evolution of the median $(u-g)_{0}$ color of blue galaxies as a 
function of the redshift. 
The small green circles show the median colors of the blue galaxies for each 
cluster candidate. 
The large red circles and the blue diamonds are the median $(u-g)_{0}$ values 
of blue galaxies in cluster (red circle) and in field (blue diamond). 
Vertical error bars show the first and the third quartiles, and red horizontal 
error bars show the range of redshift of individual clusters included at 
each redshift bin.
The magenta circle shows the SDSS value.
The lines show the predicted evolution of $(u-g)_{0}$ colors from BC03 with 
varying values of $\tau$ and $E(B-V)$.
The green dotted lines represent the color evolution with $\tau = 1.5$ Gyr and 
$z_{f} = 5$, where $z_{f}$ is the formation redshift. The blue dashed lines 
are for the color evolution with $\tau = 2.0$ Gyr and $z_{f} = 5$, and the red 
solid lines are for $\tau = 3.0$ Gyr and $z_{f} = 8$.
When $z_f = 5$ (blue and green lines), the age ($t$) is 3.0 Gyr and 4.6 
Gyr at $z=1.5$ and $z=1$. 
If $z_f = 8$ (red lines), the ages are 3.6 and 5.1 Gyr at $z=1.5$ and 1.0.
Three lines with same color and style represent the color evolution of 
templates with $E(B-V) = 0.0$, 0.1, and 0.2 from bottom to top. 
Solar metallicity is assumed in the case of the synthetic color evolution. 
This reddening of blue galaxy color with decreasing redshift indicates 
the average ageing of stellar populations in blue galaxies. \label{bcolev}}
\end{figure}

In Figure~\ref{bcolev}, we show the redshift evolution of the median color of 
blue galaxies, both for clusters and field. 
In this figure, the small green circles show the median color of blue galaxies 
and the redshift for each cluster candidate, and the red, large circles show the 
median colors and the median redshifts of 
blue cluster galaxies summed in each redshift bin. 
The blue diamonds represent the median values of field blue galaxies at each 
redshift bin. 
The corresponding SDSS value, which is derived from the MPA-JHU 
catalog\footnote{http://www.mpa-garching.mpg.de/SDSS/}, is shown as 
the magenta circle.
The expected color evolution of galaxies from the BC03 model with various (delayed) 
SFHs are also shown as the green dotted lines ($\tau = 1.5$ Gyr, 
$z_{f} = 5$), blue dashed lines ($\tau = 2.0$ Gyr, $z_{f} = 5$), and red 
solid lines ($\tau = 3.0$ Gyr, $z_{f} = 8$).
Here, $z_f$ is the formation redshift. 
For each color, the lines from the bottom to the top are with the increasing 
amount of dust attenuation from $E(B-V) = 0$ with the increment with 0.1.

As already mentioned, the blue peak evolves to become redder with decreasing 
redshift, reflecting the average aging of the stellar population in the blue galaxies. 
Apparently, the BC03 models with $\tau = 3.0$ Gyr, $z_{f} =  8$ (with $E(B-V) = 0.0$ 
and 0.1 --- i.e., upper two red solid lines) or the models with $\tau = 2.0$ Gyr, 
$z_{f} =  5$ (blue dashed curves) seem to well bracket the 
observational trend of blue galaxies. 
This does not necessarily mean that the individual galaxies would evolve 
with this SFH, even though it can be a representation of global SFH within 
this redshift range.
Interestingly, these SFHs have a peak of SFR at $z \sim 1.8$ or 2, showing a 
coincidence with the peak of the global SFR density evolution.
It should be noted that the difference between cluster- 
and field-galaxies is not significant, considering the associated ranges of 
error bars and the fact that median $(u-g)_{0}$ color can be affected by 
non-negligible fraction of green galaxies. 

In interpreting this color evolution of blue galaxies, we should 
consider the fact that the blue galaxy population is not a closed one. 
As galaxies evolve to become redder, these galaxies would move out to 
the red galaxy population. 
Also, low mass blue galaxies that were originally out of our mass-cut 
would be progressively included into the blue galaxy population as 
their stellar mass grows. 
The effect of these is to prevent the median color of the blue galaxy 
population from being reddened quickly.
Therefore, it is very likely that the SFRs of the blue galaxies would 
evolve more rapidly (i.e., having smaller value of $\tau$) than the one 
represented by the red curves in Figure~\ref{bcolev}.   
Also, we cannot rule out the possibility of the change in the median 
dust attenuation values of the blue galaxies with redshift. 
The distributions of $E(B-V)$ show little, if any, evolution within the 
redshift range, $0.5 \lesssim z \lesssim 2.0$, with a hint of slight 
decrease of the mean $E(B-V)$ at the highest redshift bins ($z > 1.6$). 
If we accept this small amount of evolution of dust attenuation, it 
means that the higher dust attenuation at lower redshift contributes, 
to some extent, to the redder color (of the blue galaxies) at lower 
redshift, requiring larger value of $\tau$. 
This will compensate, to some extent, the effects of migration of fading 
galaxies out of the blue population, making our estimation of the 
(global representative) SFH more robust. 

\begin{figure}[h]
\plotone{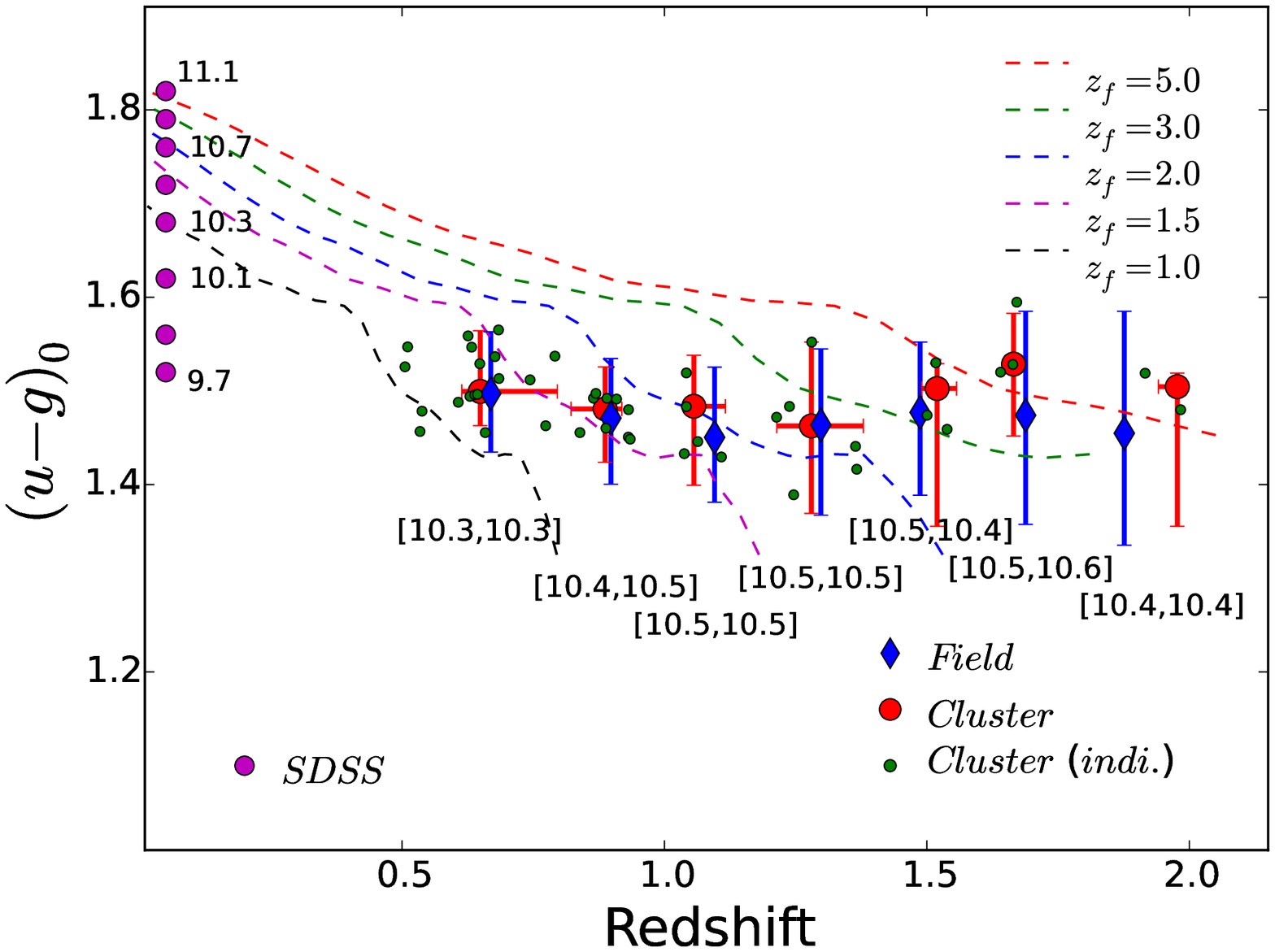}
\caption{The evolution of the median $(u-g)_{0}$ color of quiescent galaxies 
as a function of redshift. 
The symbol assignment is same as Figure~\ref{bcolev}. 
$(u-g)_{0}$ color of quiescent galaxies shows little evolution within the 
redshift range, $0.5 < z < 2.0$.
This implies that newly quenched galaxies keep joining the quiescent galaxy 
population within this redshift range. 
Dashed curves are the evolutionary tracks of quiescent galaxy templates from BC03 
with various formation redshift ($z_f$) as indicated in the figure. 
Purple circles are the corresponding colors of SDSS galaxies with different stellar 
masses, which increase from bottom (log$(M_{*}/M_{\odot}) \sim 9.7$) to top 
(log$(M_{*}/M_{\odot}) \sim 11.1$) with a decrement of 0.2. 
Bracketed numbers are the median stellar masses at corresponding redshifts 
for field and cluster galaxies, respectively. \label{pscolev}}
\end{figure}

The evolution of the $(u-g)_{0}$ color of quiescent galaxies is 
shown in Figure~\ref{pscolev}. 
Unlike blue galaxies, the quiescent galaxy population show little 
color evolution within the redshift range of $0.5 < z < 2.0$ either in 
clusters or in field. 
This no-evolution of $(u-g)_{0}$ color reflects the fact that $new$ 
quiescent galaxies (with the colors bluer than the $existing$ quiescent 
ones) keep joining the quiescent population while the color of the 
$existing$ quiescent galaxies becomes redder with time.
The dashed curves with various colors in Figure~\ref{pscolev} are the 
evolutionary path of BC03 quiescent galaxies with various formation 
redshift. 
These curves are shown to guide how the $(u-g)_{0}$ colors would evolve 
if no additional quiescent galaxies with bluer color join the existing 
quiescent population. 
Purple circles are the corresponding SDSS colors in various stellar mass 
bins --- log$(M_{*}/M_{\odot}) = 9.7$ for the bottom one to 
log$(M_{*}/M_{\odot}) = 11.1$ for the top one with an increment of 0.2 
in logarithmic scale.
Analysing red galaxy spectra at $z \sim 0.9$ from the Deep Extragalactic 
Evolutionary Probe 2 (DEEP2), \citet{sch06} found 
relatively young ages ($\sim 1$ Gyr) for these red galaxies with 
little or no emission lines. 
One of their interpretation is that blue SF galaxies keep joining this 
red population continuously \citep[see also][]{har06,fab07} maintaining 
their mean stellar population ages young.

\subsubsection{Red Star-forming Galaxies}\label{redsf}

\begin{figure}[h]
\plotone{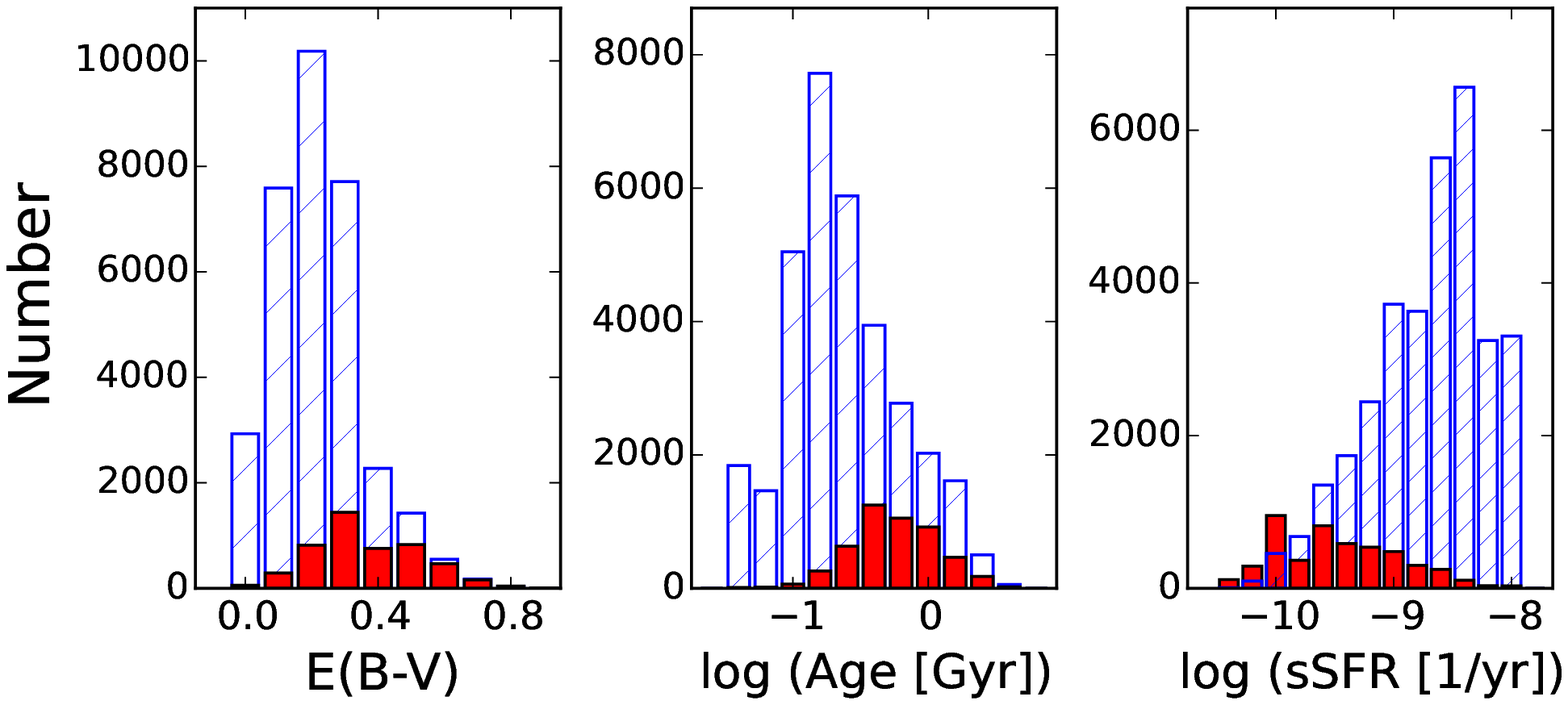}
\caption{Distributions of dust extinction (E(B-V); {\bf Left}), 
mean stellar population age ({\bf Middle}), and sSFR 
({\bf Right}) estimated from SED-fitting for blue SF (blue hatched histogram) 
and red SF (red filled histogram) galaxies.  
On average, red SF galaxies are more dustier as well as older with lower 
sSFRs. \label{redsfdst}}
\end{figure}

The $(u-g)_{0}$ color and sSFR of galaxies show a broad correlation in a 
sense that redder galaxies have in general lower sSFR. 
However, there are a non-negligible fraction of galaxies 
whose $(u-g)_{0}$ color is red ($> 1.2$) but are still forming stars actively. 
As shown in Figure~\ref{redsfdst}, these red star-forming (SF) galaxies are, 
on average, dustier, older, and have lower sSFR than blue SF galaxies.
No red SF galaxy has sSFR $\gtrsim 10^{-8}$ yr$^{-1}$ to be classified as 
`starbursts'.
These indicate that red SF galaxies are in the $fading$ stage, and migrating into 
the red quiescent population.

\begin{figure}[h]
\plotone{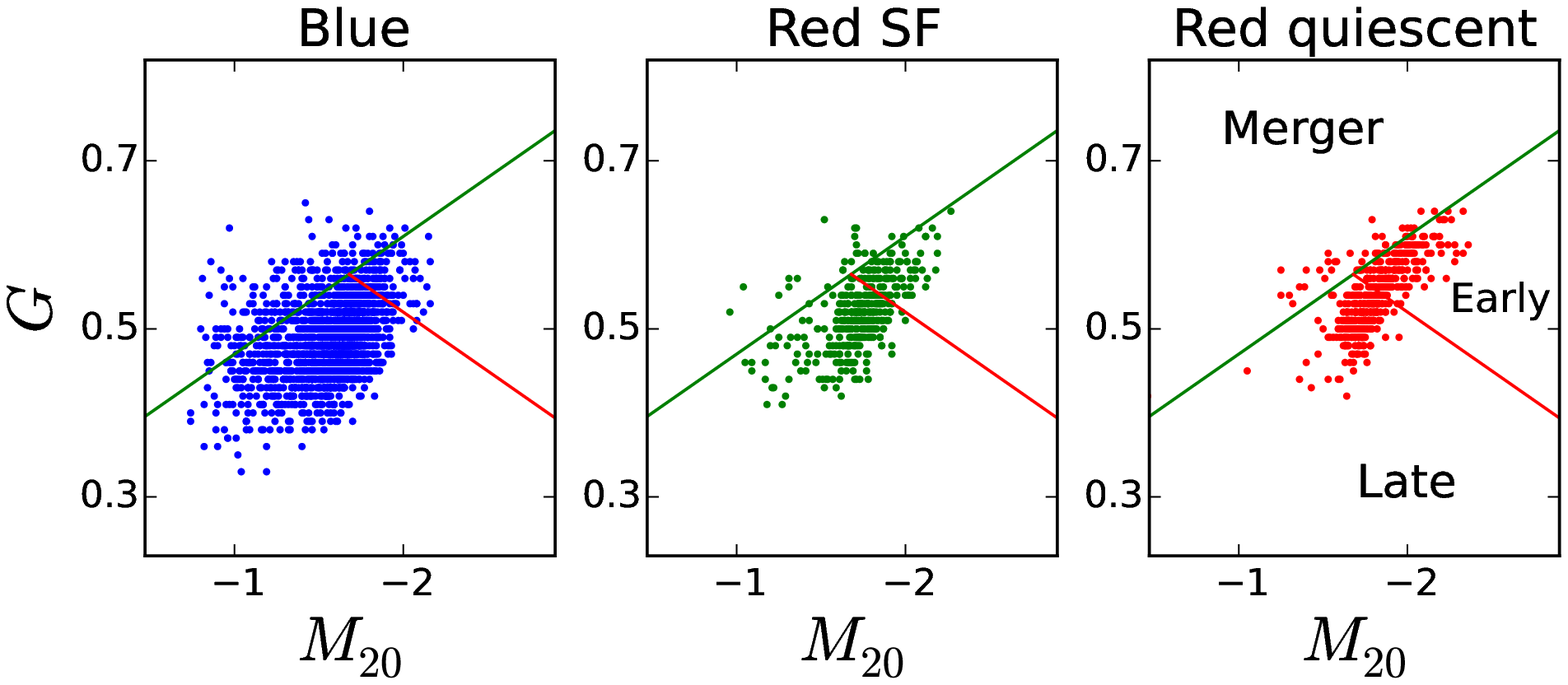}
\caption{The $G$-$M_{20}$ diagrams of galaxies in the sub-region 
of the UDS covered by the CANDELS. 
{\bf Left column} (blue points): Blue galaxies. 
{\bf Middle column} (green points): Red SF galaxies. 
{\bf Right column} (red points): Red quiescent galaxies. 
The dividing lines are from \citet{lot08}. 
Majority of red SF galaxies lie in the region between blue galaxies 
and red quiescent ones in this diagram, and only a few red SF galaxies 
are merger candidates. \label{gm20}}
\end{figure}

We investigate the morphological properties of red SF 
galaxies and compare them with the other galaxy populations to see how many red 
SF galaxies could be starburst galaxies resulting from gas-rich (wet) merger.
A portion of the UDS region ($\sim 22.3 \arcmin \times 9 \arcmin$) is covered by 
the CANDELS 
\citep[Cosmic Assembly Near-infrared Deep Extragalactic Legacy Survey;][]{gro11,koe11} 
program, and observed with the WFC3 (Wide Field Camera 3) on board the $HST$.
From the deep WFC3 H-band (F160W) image, which reaches to the 5-$\sigma$ 
magnitude limit of 27.45 with FWHM of $0.2 \arcsec$ \citep{gal13}, 
the Gini coefficients, $G$, as well as the $M_{20}$ parameters of the galaxies 
are measured. 
Figure~\ref{gm20} shows the distributions of the Gini 
coefficients and the $M_{20}$ values of the blue, the red SF, 
and the red quiescent galaxies. 
The dividing lines between the early-, late-type and major mergers are 
from \citet{lot08}. 
There are only a few merger candidates 
among red SF galaxies (in the middle panels, represented as green symbols) 
and most of the red SF galaxies lie around the boundary between 
blue (left panels) and red quiescent galaxies (right panels), supporting 
the idea that these red SF galaxies are in transition phase from blue galaxies 
to red quiescent ones.

\begin{figure*}[h]
\includegraphics[scale=0.25]{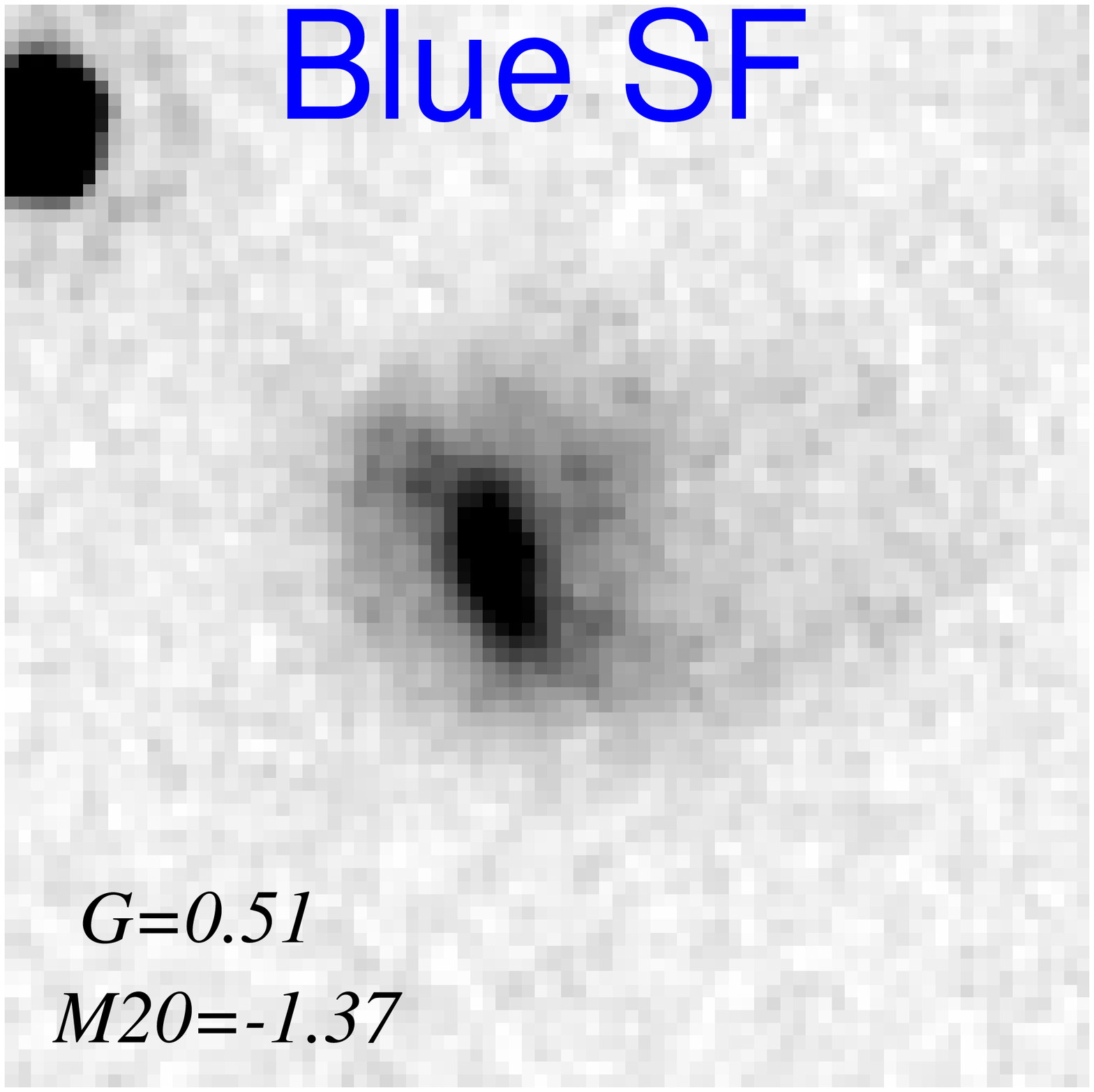}
\includegraphics[scale=0.25]{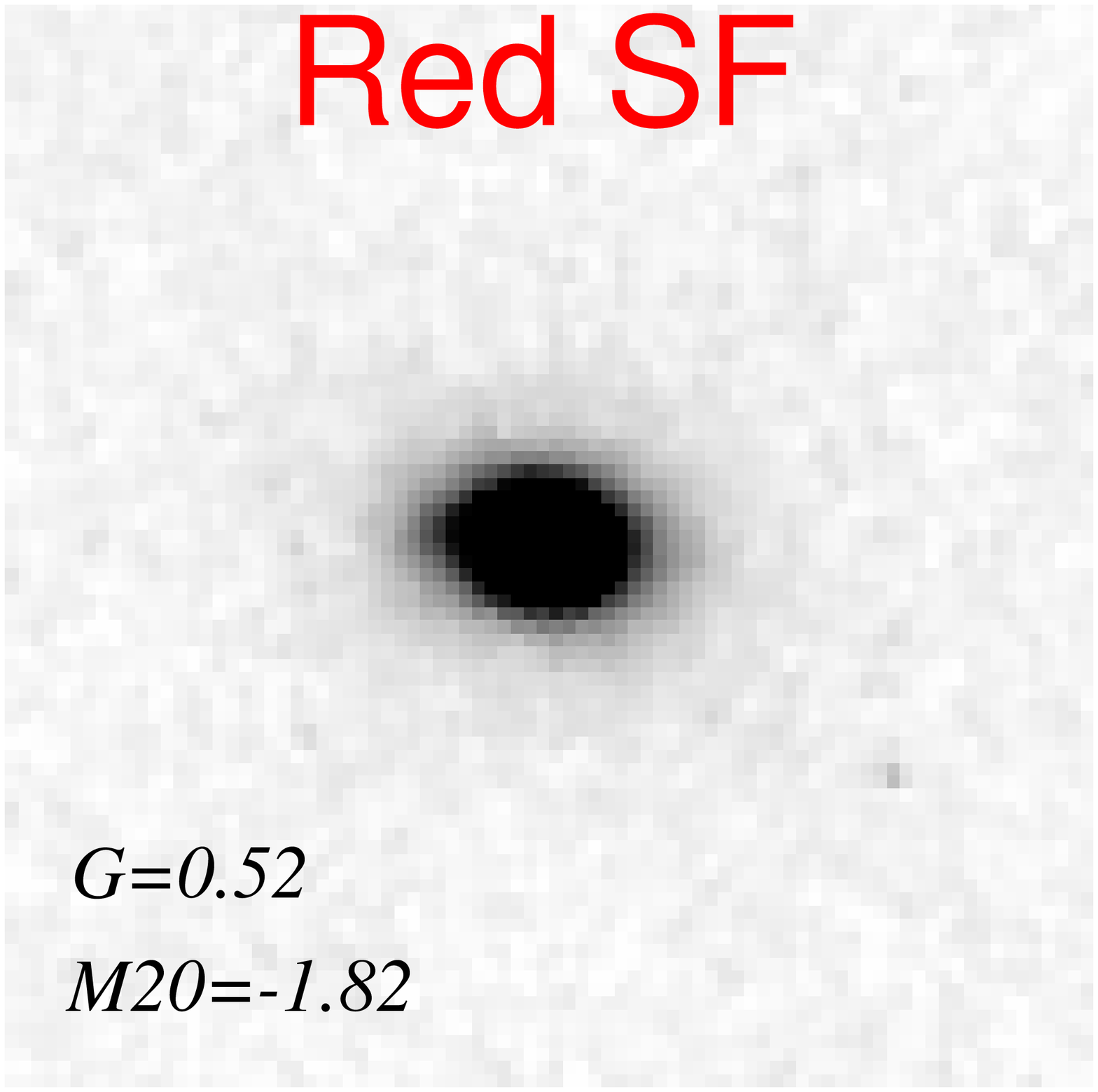}
\includegraphics[scale=0.25]{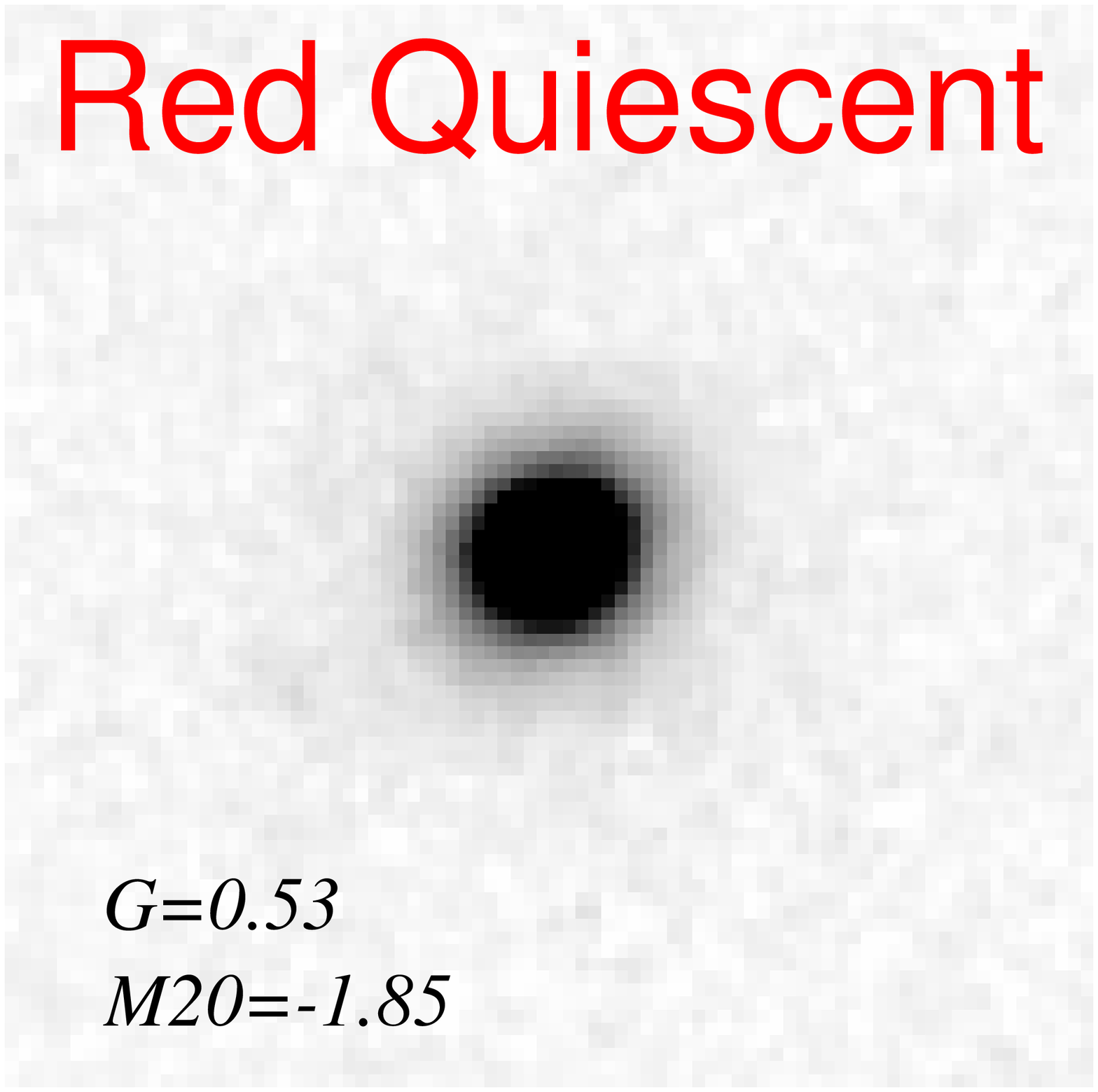}\\
\includegraphics[scale=0.25]{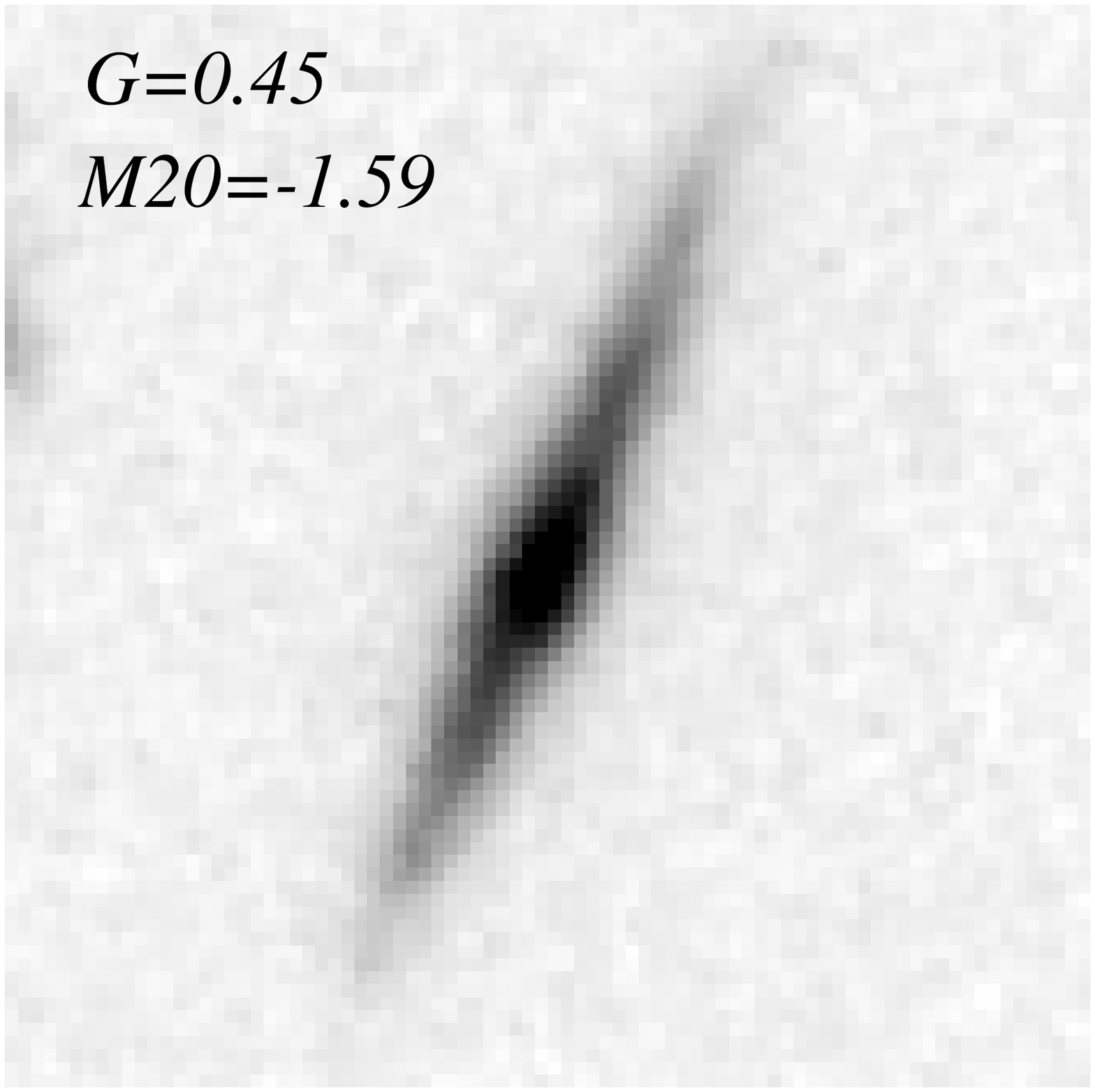}
\includegraphics[scale=0.25]{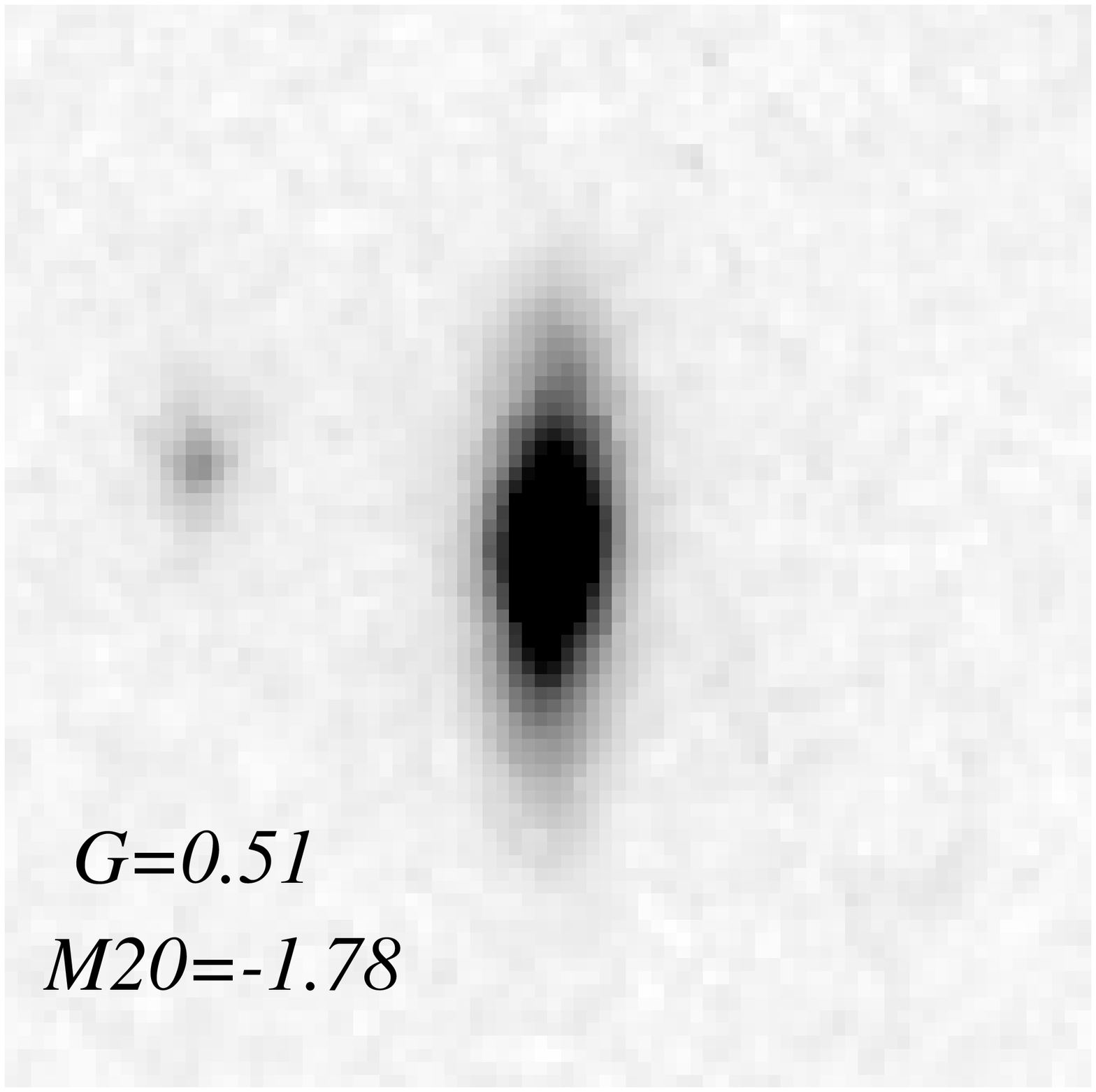}
\includegraphics[scale=0.25]{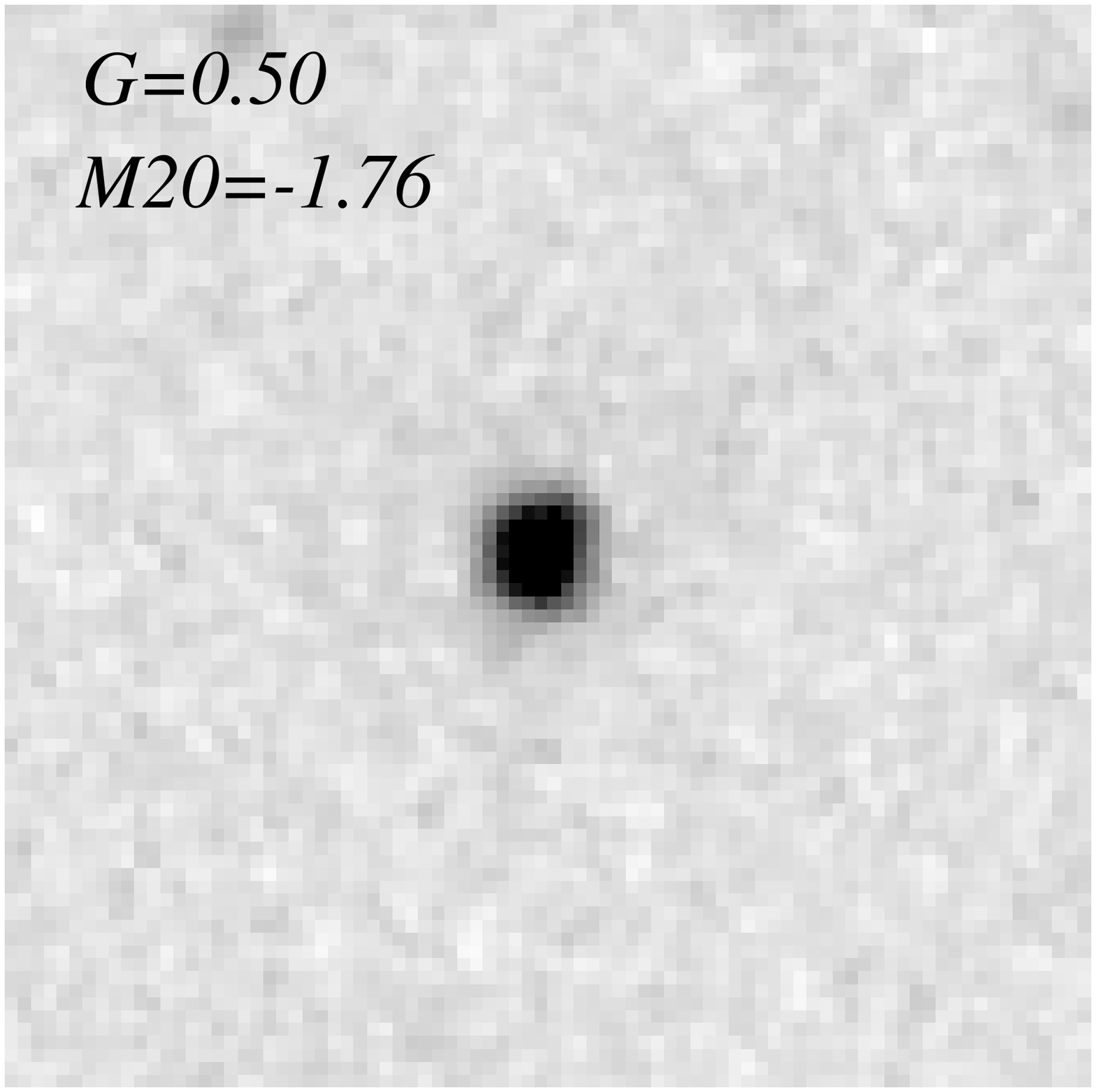}\\
\includegraphics[scale=0.25]{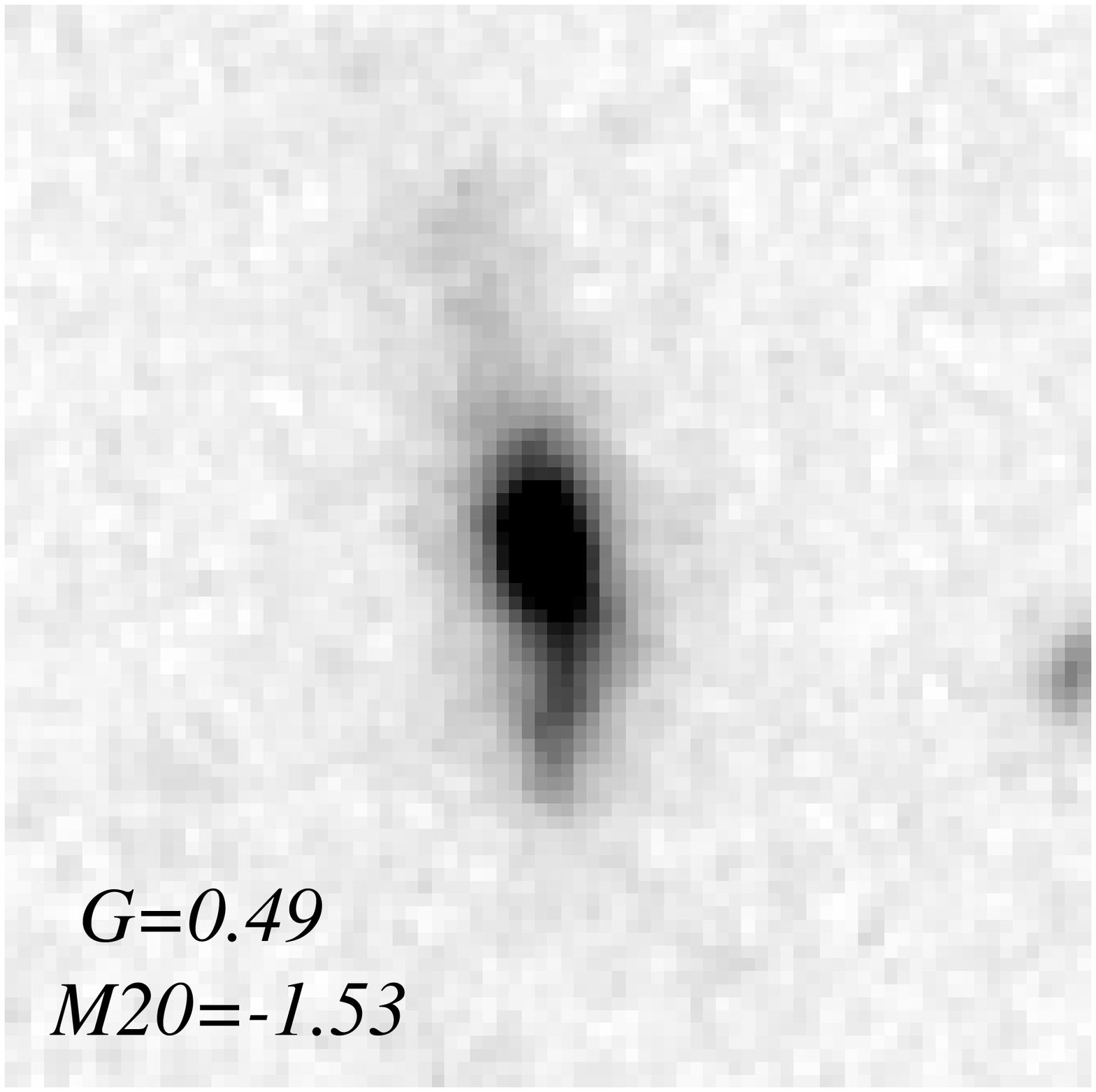}
\includegraphics[scale=0.25]{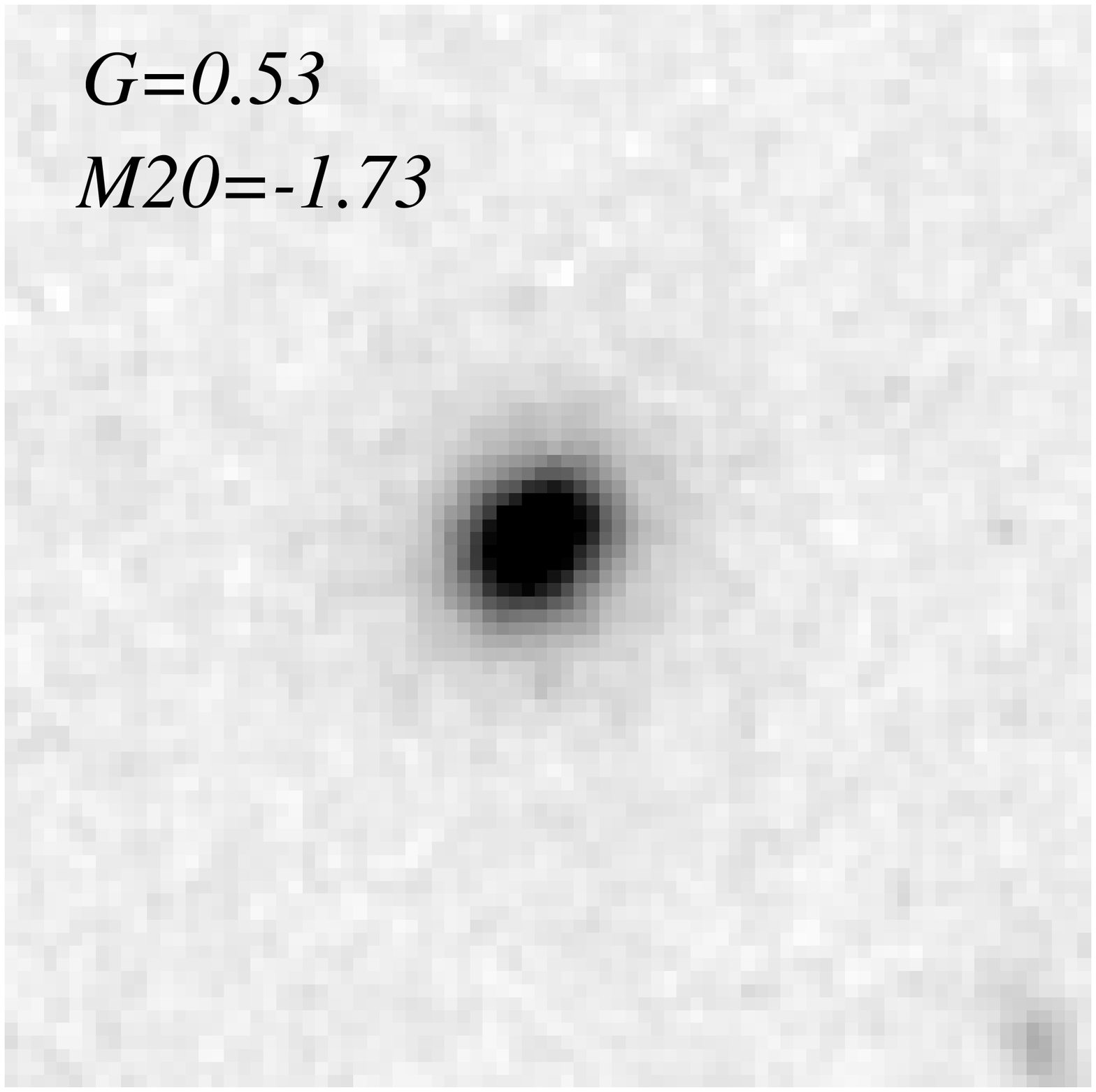}
\includegraphics[scale=0.25]{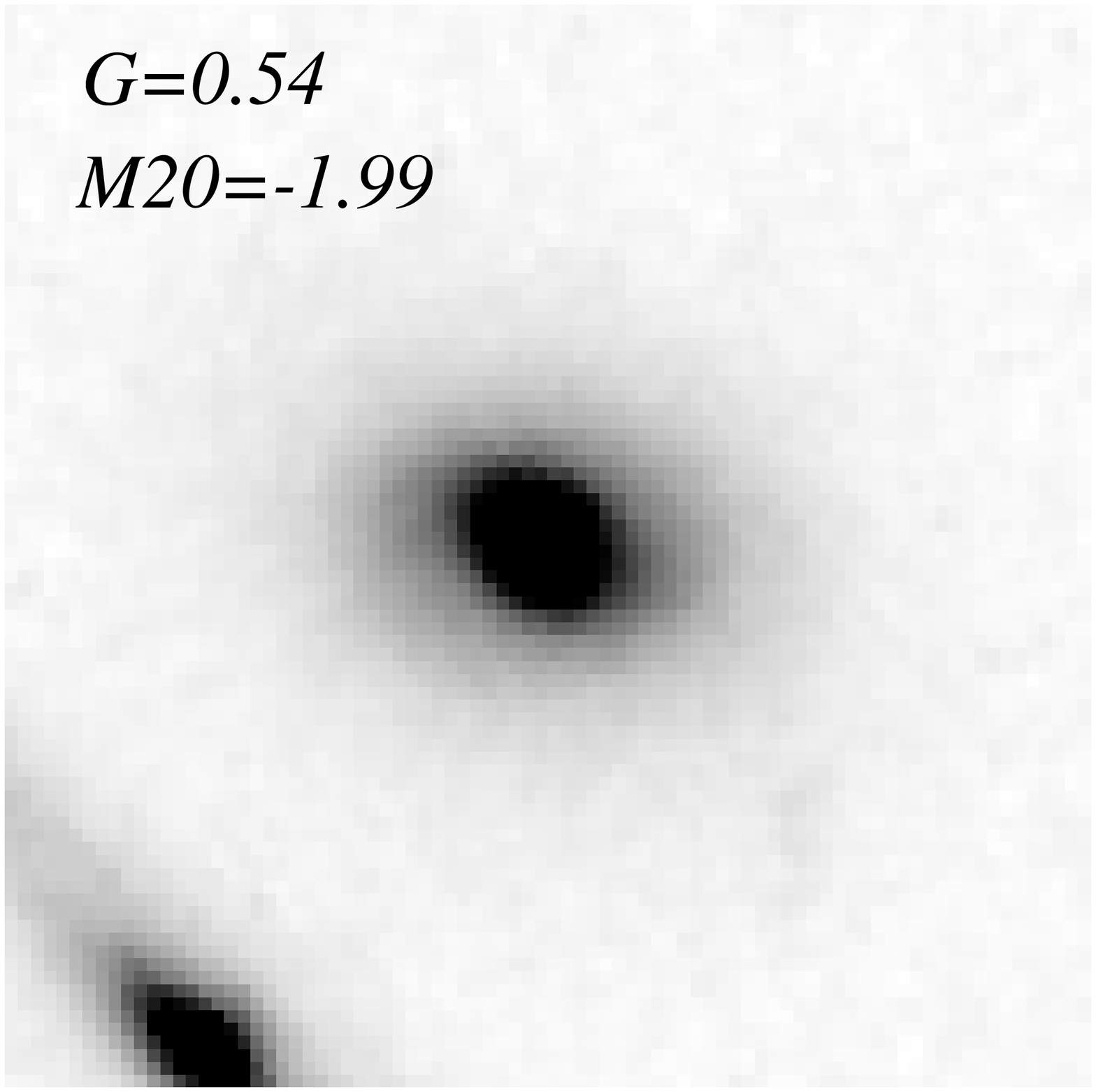}
\caption{$HST$/WFC3 $H$-band (F160W) images of cluster galaxies at $z \sim 0.65$. 
{\bf(Left)} Examples of blue star-forming galaxies. {\bf(Middle)} Red 
star-forming galaxies. {\bf(Right)} Red quiescent galaxies. 
Each image cut size is $\sim 5\arcsec \times 5\arcsec$. 
We show the values of Gini coefficient ($G$) and $M20$ of each galaxy as well. \label{himage}}
\end{figure*}

We also visually inspect the $HST$/WFC3 images of our galaxies. 
The blue SF population mostly consists of extended objects with 
disk-like morphology with small fraction of disturbed ones hinting the recent 
experience of merger or interaction. 
On the other hand, red SF galaxies are mixture of disky, spherical, 
and compact objects, reinforcing that these objects are in transition from 
blue galaxies into red quiescent ones. 
In Figure~\ref{himage}, we show example $HST$/WFC3 $H$-band images of cluster 
galaxies at $z \sim 0.65$.  

\subsection{SF properties and Evolution of Cluster and Field Galaxies}\label{sfnevol}

In previous section, we investigated the rest-frame optical color properties 
of galaxies in clusters and field. 
Now, in this section, we concentrate on the SF properties of galaxies and 
analyse how the SF properties of galaxies within the high-redshift clusters 
evolve with time and also compare this with that of the field galaxies in 
the same redshift range. 

\begin{figure}[h]
\plotone{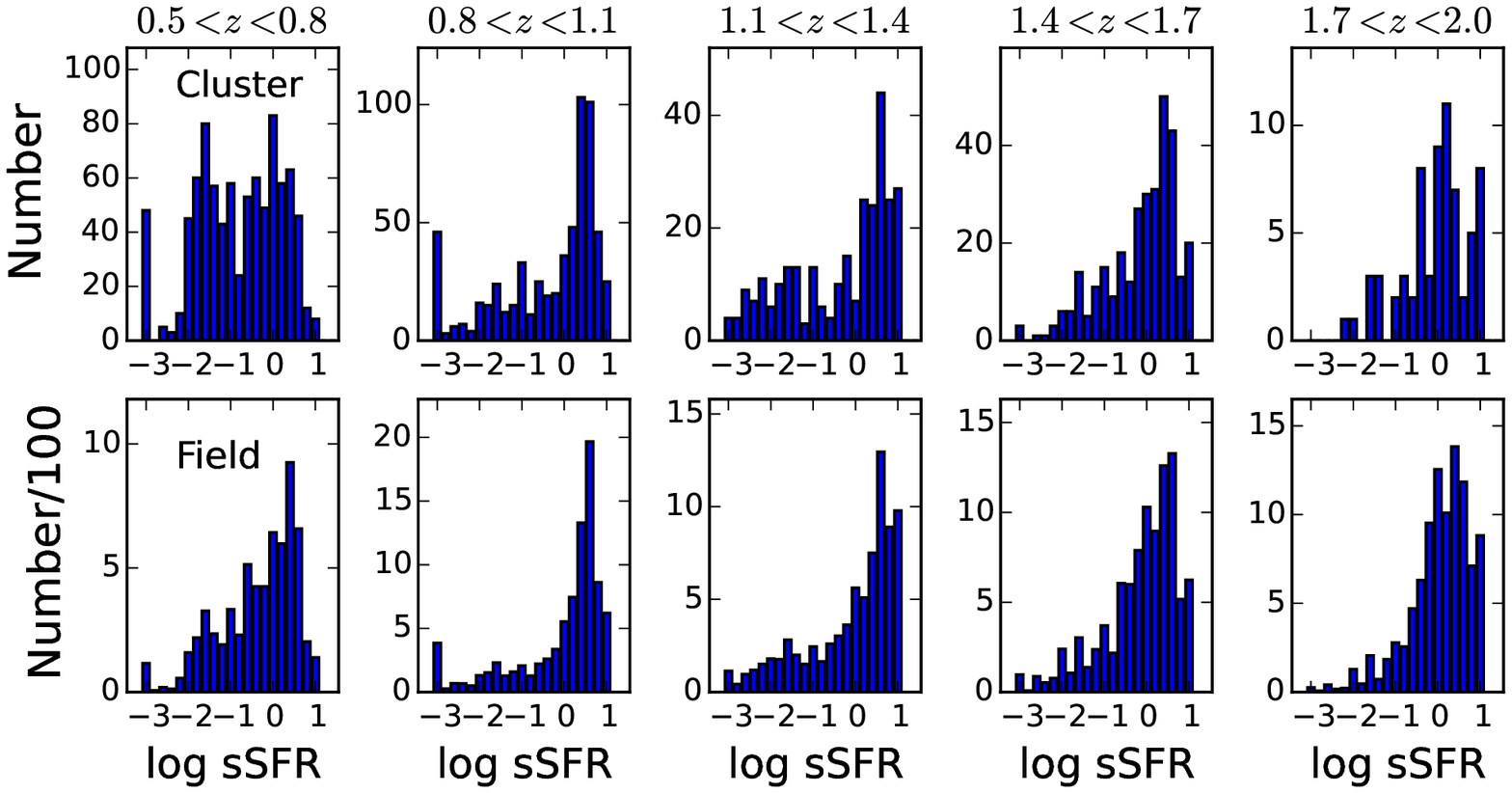}
\caption{Top: The sSFR distributions of cluster galaxies in five 
redshift bins. The cluster galaxies are summed in each redshift bin. Bottom: The 
sSFR distributions of the field galaxies in the same redshift bins 
as in the top row. The sSFR is given in the unit of Gyr$^{-1}$. 
In each panel, we assign sSFR = $10^{-3}$ Gyr$^{-1}$ to all quiescent 
galaxies with sSFR < $10^{-3}$ Gyr$^{-1}$.
The sSFR distribution of cluster SF galaxies shows clear difference from that of 
field SF galaxies only at the lowest redshift bin ($z \sim 0.65$), while relative 
fraction of quiescent galaxies shows difference between clusters and field at 
$z \lesssim 1.4$. \label{ssfrdstclfl}}
\end{figure}

In Figure~\ref{ssfrdstclfl}, we compare the distributions of 
sSFR of the cluster- and the field galaxies at five redshift bins 
with the same redshift binning as in the Figure~\ref{coldstclfl}.
In this figure, we assign sSFR = $10^{-12}$ yr$^{-1}$ to all quiescent 
galaxies with sSFR $< 10^{-12}$ yr$^{-1}$.
Similarly with color distribution, the sSFR distribution of SF galaxies 
in clusters and in the field are similar at high redshift bins ($z \geq 1.4$), 
while it shows a clear difference from that of the field galaxies at 
the lowest redshift bin ($z \leq 0.8$). 
For the quiescent galaxy fraction (over SF ones), the difference 
between clusters and field starts to show up from $z \sim 1.25$. 
While the evolutionary trends of sSFR distributions is similar with that 
of the color distributions shown in Figure~\ref{coldstclfl}, the sSFR 
distribution changes more slowly (especially for the cluster galaxies) 
than the color distribution. 
For example, at $z \sim 0.95$, the color distributions are already 
different between cluster and field, while the sSFR distributions of 
SF galaxies still remain similar in field- and cluster-environments. 

\begin{figure}[h]
\plotone{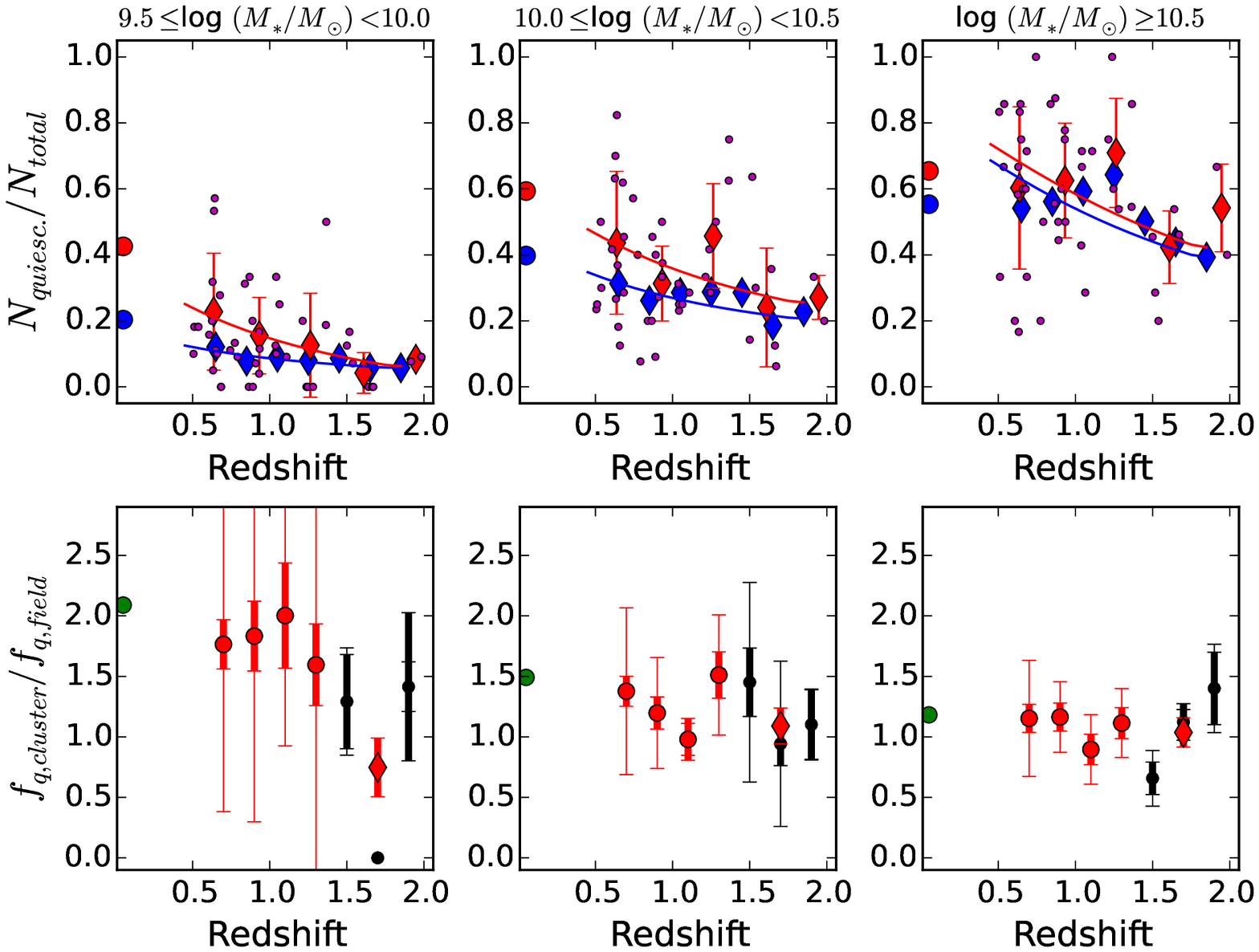}
\caption{$\bf{Upper}$: The quiescent fraction evolution of galaxies in three 
stellar-mass bins (log $M_{*}/M_{\odot} \sim 9.75$ (left), 
10.25 (middle), and $\geq 10.5$ (right). 
The purple dots show the quiescent galaxy fraction of the individual cluster candidate 
and the red diamonds with the error bar are the number-weighted mean and the standard 
deviation of cluster galaxies summed at discrete redshift bins ($\Delta z = 0.3$).
The blue diamonds are the quiescent fraction of the field galaxies at each 
redshift bin ($\Delta z = 0.2$). 
The red and blue circles in each panel show the cluster and field quiescent fraction at 
each corresponding stellar mass of the SDSS galaxies from \citet{bal06}. 
The difference of quiescent fraction in different stellar mass bins is greater 
than the difference between different environment in the same stellar mass bin. 
Solid curves in each panel show the best-fit quiescent fraction evolution 
(Equation 4) for cluster (red) and field (blue) galaxies, respectively. 
$\bf{Lower}$: The excess of the quiescent fraction in the clusters over the field value, 
defined as $f_{q,cluser}/f_{q,field}$, where $f_{q,cluster}$ 
and $f_{q,field}$ are quiescent fraction in the clusters and in the field, respectively.
The thick error bars show the standard deviation at each redshift bin, while the thin 
error bars reflect the spread among clusters in the given redshift bin. 
We observe this excess becomes significant ($>1.5$) only in the smallest mass bin 
(lower left panel) at $z < 1.2$. \label{psfracmsbin}}
\end{figure}

Next, we compare the evolution of the quiescent galaxy fraction in 
cluster and in field in Figure~\ref{psfracmsbin}, dividing the sample 
into three stellar-mass bins : $9.5 \leq$ log $M_{*}/M_{\odot}$ $< 10.0$ 
(left panel), $10.0 \leq$ log $M_{*}/M_{\odot}$ $< 10.5$ (middle), and 
log $M_{*}/M_{\odot}$ $\geq 10.5$ (right). 
In this figure, the filled blue diamonds represent the quiescent 
fraction of field galaxies at each redshift bin, while 
the quiescent fraction of individual clusters is shown as magenta dots. 
At several redshift bins, we sum the cluster galaxies in each $z$-bin, 
and calculate the mean and the standard deviation.
This is shown as the red diamonds with error bar in the figure.
The red and blue circles are for the SDSS (Sloan Digital Sky 
Survey) galaxies from \citet{bal06}. 
We derive these values from their Equation (9), and choose the 
values with $\sigma = -0.3$ as field values and $\sigma = 0.9$ as 
cluster values in that equation. 
And, we apply the correction for red SF galaxies based on 
\citet{hai08}.

From Figure~\ref{psfracmsbin}, we find several interesting aspects in the evolution 
of quiescent galaxy fraction.
First, the increase of the quiescent fraction slows down 
from the redshift $z \sim 1.3-1.4$ for galaxies with their stellar mass, 
log $(M_{*}/M_{\odot}) \geq 10$, more significantly for the field galaxies, 
while it evolves fast from $z \sim 2$ down to $z \sim 1.3$ for massive 
galaxies ($M_{*} \geq 10^{10.5} M_{\odot}$).
At the two high stellar-mass bins, the quiescent fraction remains nearly 
unchanged from $z \sim 1.3-1.4$ both in field and in cluster.
Comparison with the local values (the red and the blue circles) shows 
that the quiescent fraction has already reached to the local value for 
the most massive galaxies (log $(M_{*}/M_{\odot}) \geq 10.5$) at 
$z \sim 0.6$ both in clusters and in field (the right panel), while the 
growth of quiescent fraction must be accelerated at $z \lesssim 0.5$ to 
match the SDSS values in the case of low mass galaxies with 
log $(M_{*}/M_{\odot}) < 10$. 
This near-constant quiescent fraction of massive galaxies is in good 
agreement with the previous studies \citep[e.g.,][]{imm02,ilb13}.
This change in the increase of quiescent fraction 
indicates that the redshift range $z \gtrsim 1.3$ defines the era of the 
rapid build-up of massive quiescent galaxies, which is in broad agreement 
with the finding of actively star-forming galaxy clusters at $z > 1.3$ 
by several authors (e.g., Tran et al. 2010; Zeimann et al. 2012; 
Santos et al. 2014).

Second, the fraction of quiescent galaxies is higher in clusters than 
in field at $z < 1.4$ for galaxies with $M_{*} < 10^{10.5} M_{\odot}$, 
even though the scatter among the individual cluster candidates 
is quite large.
From this, we can speculate that the environmental quenching becomes 
to work more strongly at redshift, $z \lesssim 1.4$, but with a certain 
amount of cluster-to-cluster variation. 
Also, we can see that the difference in the quiescent galaxy fraction 
between the cluster and the field environments is more significant 
for low-mass (log $(M_{*}/M_{\odot}) < 10$) galaxies (the left panel) 
at redshift $z \lesssim 1$. 

Last, as can be seen clearly in this figure, the quiescent fraction is 
a strong function of the stellar mass rather than their environment. 
At the highest mass bin, the $cross$-$over$, which we define as the 
redshift or the epoch when the quiescent fraction starts to exceed 
$50~\%$, occurs already at $z \sim 1.5$.
In sharp contrast, in the lowest mass bin, the quiescent fraction 
never reaches the $cross$-$over$ down to $z \sim 0.5$ both in field and 
in cluster.
Actually, the quiescent fraction is $\lesssim 0.2$ throughout the 
redshift range in field.
This strong mass-dependence of quiescent fraction evolution well agrees 
with the recent results of \citet{mou13} --- who found a nearly 
constant number density of massive quiescent galaxies at 
$z \lesssim 1$ while the corresponding value rises rapidly with 
decreasing redshift for less massive galaxies, and of \citet{hua13}.
This strong mass dependence and the weaker environmental dependence 
dictates that the stellar mass plays a more dominant role in shaping the 
SFH of galaxies than the cluster-specific processes. 

To gain an insight how rapidly (or how slowly) star formation is quenched 
for galaxies with different stellar masses, we devise a simple model 
explaining the evolution of star-forming galaxy fraction, $f_{sf} (t)$, 
as follows.

\begin{equation}\label{sffevol}
f_{sf} (t) = f_{sf} (0) \times (\onehalf)^{t/\gamma}, 
\end{equation}

where, $f_{sf} (0)$ is the fraction of SF galaxies at an initial time 
$t = 0$. 
In this model, $\gamma$ is a $half$-$life$ of SF galaxy population, which 
means that the fraction of SF galaxies becomes half of its initial values 
after the time $t = \gamma$. 
This simplified model assumes that the total number of galaxies does not 
change with time --- i.e., no galaxy is added to or removed from the sample, 
and no merger occurs between galaxies. 

Then, the quiescent fraction, $f_{quies.}$ would increase as, 

\begin{equation}\label{psfevol}
f_{quies.} (t) = 1 - f_{sf} (0) \times (\onehalf)^{t/\gamma}.
\end{equation}

We fit our data points to this simple model, and we show this 
$f_{quies.}$ for the best-fit values of $\gamma$'s as solid 
curves in each panel in the upper row of Figure~\ref{psfracmsbin}. 
In each panel, the red and the blue curves are for cluster and field 
galaxies, respectively. 
For cluster galaxies at $0.5 < z < 2.0$ (solid red curves), 
the best-fit values for $\gamma$ are $15 \pm 2.2$, $9.7 \pm 3.6$ 
and $4.4 \pm 1.9$ Gyr for the lowest, middle and highest stellar-mass bins. 
In the case of field galaxies (solid blue curves), the corresponding values 
are $46 \pm 7.1$, $18 \pm 3.9$, and $5.2 \pm 1.4$ Gyr, each. 
This again demonstrates that: (1) massive galaxies become quiescent 
more rapidly both in clusters and in field (i.e., having smaller 
value of $\gamma$) than less massive galaxies, and 
(2) star formation is more rapidly quenched in clusters than in field.

\begin{figure}[h]
\plotone{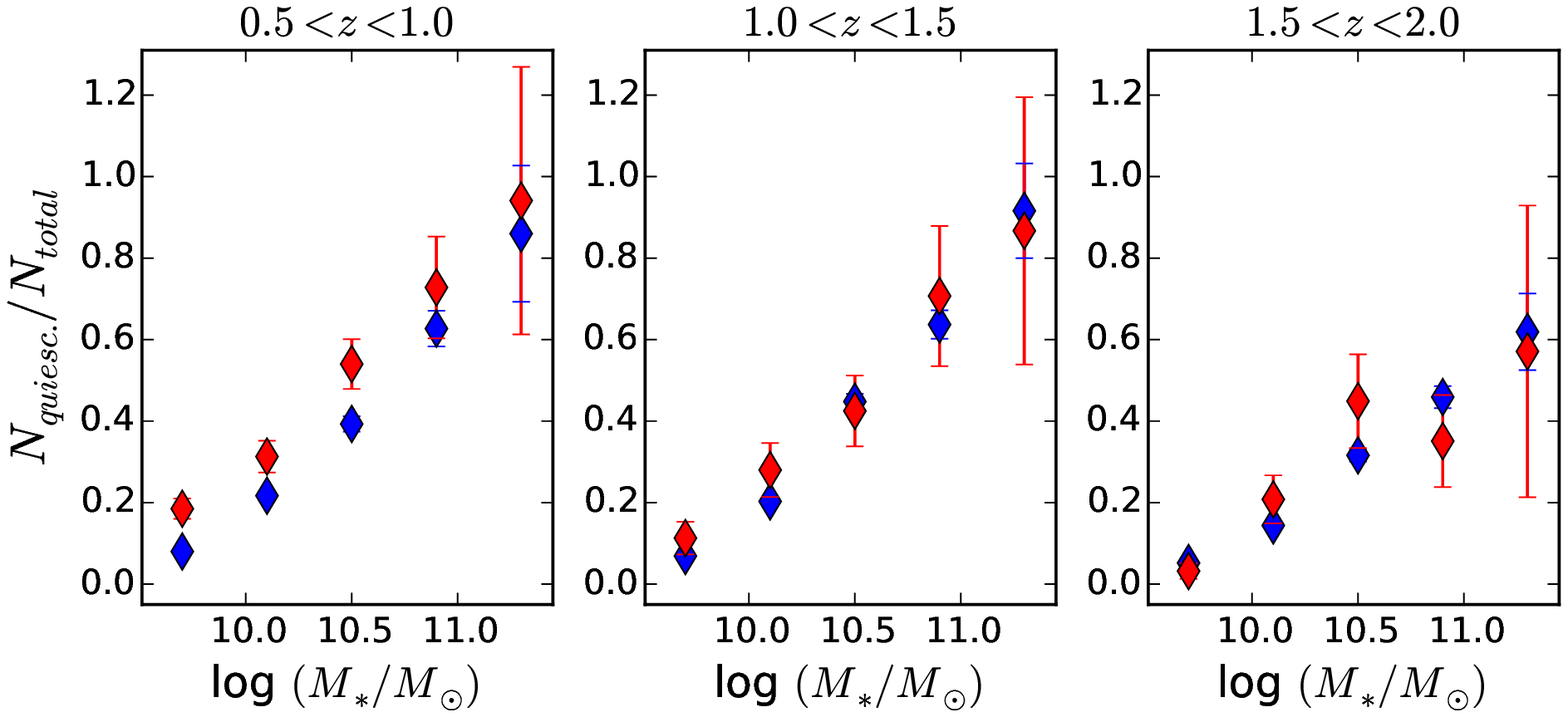}
\caption{The quiescent fraction evolution of galaxies in three redshift bins 
($z \sim 0.75$ (left), 1.25 (middle), and 1.75 (right). 
The red and blue diamonds with the error bar are the number-weighted mean and the standard 
deviation of cluster and field galaxies summed at discrete stellar-mass bins, respectively.
The clear dependence of quiescent fraction on the stellar-mass can be seen.\label{psfraczbin}} 
\end{figure}

The lower panels of Figure~\ref{psfracmsbin} show the ratio of 
the quiescent fraction in clusters 
($f_{q,clustesr}$) to that in the field ($f_{q,field}$).
Here, we take an average value of the highest three redshift bins (shown 
as the black circles)
and show it as a red diamond in each panel.
In the two high mass bins (middle and right panel), the fraction is always 
$\lesssim 1.5$ with little evolution.
However, this fraction increases rapidly with decreasing redshift for low 
mass galaxies (log $M_{*}/M_{\odot}$ $< 10$), and is higher than that in 
the higher mass bins at $z < 1.4$.
This also indicates that the environmental quenching becomes more 
significant for low mass galaxies at $z < 1.4$.

To see the effect of stellar mass in determining the quiescent fraction 
more clearly, we show the stellar-mass dependent quiescent fraction at three 
redshift bins in Figure~\ref{psfraczbin}. 
In this figure, we can clearly see the strong stellar-mass dependence of 
the quiescent galaxy fraction at all three redshift bins 
as well as a clear excess in the quiescent galaxy fraction in clusters compared to 
the field for low-mass galaxies (log $(M_{*}/M_{\odot}) \lesssim 10.5$) at 
the lowest redshift bin (left panel).

Summarizing, our investigation of the quiescent galaxy fraction and its 
evolution reveals several important 
aspects about quiescent galaxy formation: 
(1) At $z > 1.3$, quiescent, massive galaxies were built up rapidly, with 
a similar rate in both clusters and field. 
By z ~ 1.3, most of the massive,quiescent galaxies were built up. 
(2) At $z < 1.3$, the environmental dependence of quiescent fraction become 
visible. The difference between cluster- and field-environment is clearer 
for low-mass (log $(M_{*}/M_{\odot}) < 10.0$) galaxies. 
(3) Stellar mass plays more dominant role in determining the 
quiescent fraction than the environment throughout the entire redshift range.

\subsection{Quenching Efficiency}

\begin{figure}[h]
\plotone{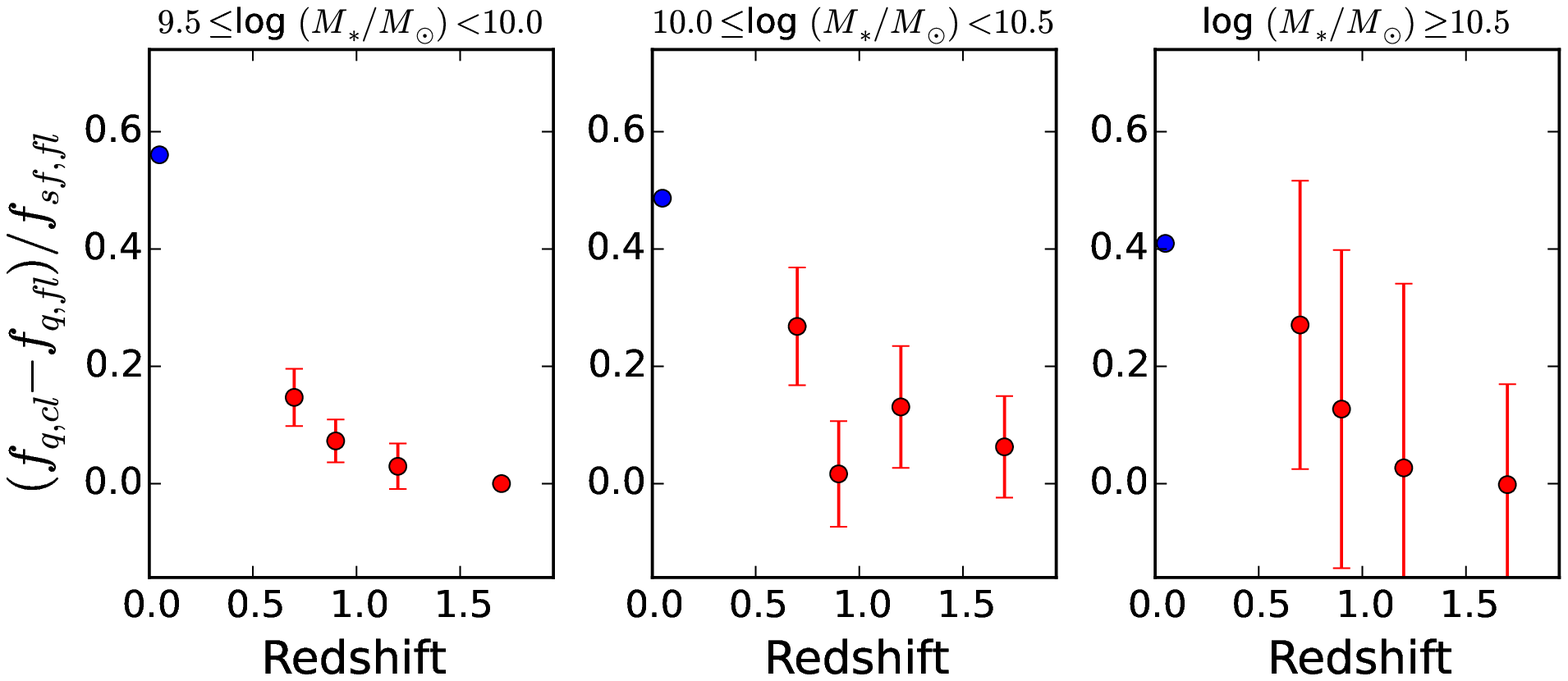}
\caption{Evolution of environmental quenching efficiency, defined as the 
excess of quiescent galaxy fraction in clusters over field divided by SF galaxy 
fraction in field. Three panels show this evolution in different stellar mass bins. 
Red circles are our results and blue circles are SDSS values. 
There is no clear difference between different stellar mass bins, except large 
scatter in the highest mass bin (log $(M_{*}/M_{\odot}) \geq 10.5$; right panel). 
\label{psdiffevol}}
\end{figure}

In the previous section, we have shown that the quiescent galaxy fraction shows 
a clear stellar-mass--dependent trend, in a sense that the fraction of quiescent 
galaxies is higher for more massive galaxies throughout the redshift range, 
$0.5 \lesssim z \lesssim 2$ (upper panels of Figure~\ref{psfracmsbin}), and 
also that the excess of the quiescent galaxy fraction in clusters over field environment 
is higher for low-mass ($< 10^{10}$M$_{\odot}$) galaxies, especially at redshifts 
lower than $z < 1.2$ (lower panels of Figure~\ref{psfracmsbin}). 
Now, we investigate the {\it environmental quenching efficiency} 
\citep[e.g.][]{vand08,pen10,qua12}. 
We define this environmental quenching efficiency as, 

\begin{equation}
(f_{q,cluster} - f_{q,field})/f_{sf,field},
\end{equation}

where $f_{q,cluster}$ and $f_{q,field}$ are the fraction of quiescent galaxies 
in clusters and in field, respectively, and $f_{sf,field}$ is the fraction of 
star-forming galaxies in field ($=1-f_{q,field}$).
This quantity measures the 
fraction of SF galaxies in field that would have become quiescent if they were 
in cluster. 
Figure~\ref{psdiffevol} shows the evolution of the 
{\it environmental quenching efficiency} in different stellar mass bins. 
Our results shows no clear stellar-mass dependent trend --- even though 
the scatter is larger in the highest mass bin due to the small number of massive 
galaxies and the small $f_{sf,field}$ values in this mass bin --- confirming 
previous results \citep{pen10,qua12} but extending the probed redshift 
range to $z \sim 2$. 

How can this result --- no significant mass dependence in environmental quenching 
efficiency --- be reconciled with the results in the previous section (e.g., 
Figure~\ref{psfracmsbin}) --- the high excess of the quiescent fraction 
in clusters in the lowest mass bin?
This difference arises because there is another quenching mechanism --- 
i.e., {\it mass quenching}. 
Massive galaxies are affected strongly by mass quenching, and they are mostly 
quiescent already at $z \sim 1.3$, leaving little room for the environmental 
quenching to play a significant role.
On the other hand, the lower mass galaxies are quenched only mainly through 
the environment effects that becomes significant at $z < 1.3$.

\section{Summary and Conclusion}

In this paper, using deep optical to MIR data in the UKIDSS/UDS field, 
we have found 46 high-redshift galaxy clusters up to $z \lesssim 2$, among 
which 27 are newly found.
We analyse the stellar population properties, such as color and SFR, of 
galaxies both in cluster and field environments over a wide redshift 
range to understand the effects of cluster environment on the galaxy evolution. 

Through this analysis, we have found that the quiescent galaxy fraction 
increases rapidly with decreasing redshift at $z \gtrsim 1.3$, and the increase 
is slowed down at lower redshift for massive (log $(M_{*}/M{\odot}) > 10.5$) 
galaxies. This trend points to the redshift range $z \gtrsim 1.3$ as the era of 
the rapid build-up of massive quiescent galaxies --- the epoch when many 
massive galaxies stop their star formation and become quiescent. 
The difference in the quiescent galaxy fraction between clusters and 
the field increases at $z \lesssim 1.2$-1.4, which coincides with the epoch when 
the increase in the quiescent galaxy fraction begins to slow down. 
In this redshift range ($0.5 \lesssim z \lesssim 1.4$), the quiescent 
galaxy fraction remains nearly unchanged in field, while it keeps increasing 
in clusters (but more slowly than at $z \geq 1.4$). 
This difference is only significant 
for low mass galaxies with log ($M_{*}/M_{\odot}$) $\leq 10.0$.

The environmental quenching efficiency shows no clear stellar-mass dependence, 
which is in agreement with previous results \citep[e.g.,][]{pen10,qua12}, while 
our results extend this up to higher redshift ($z \sim 2$). 
However, the effects of environmental quenching appears more significantly 
for low-mass galaxies, because massive galaxies are affected by another quenching 
mechanism --- i.e., mass-quenching --- and most of these massive galaxies are
already quenched at $z > 1$.
 
At $z > 1.4$, the quiescent fraction of galaxies shows no clear dependence 
on their environment, in contrast to its strong dependence on stellar mass. 
This infers that the SFH of galaxies is mainly shaped by their stellar mass 
during the early phase of evolution.
The effects of the cluster environment on determining the quiescent fraction 
of galaxies seem to be more significant for less massive galaxies with 
$M_{*} < 10^{10}\,M_{\odot}$, for which mass quenching does not yet affect their 
SF activity, at $z < 1.4$.
The delayed appearance of the cluster-environment effects on the 
quiescent galaxy formation for low-mass galaxies implies that the 
cluster-specific processes which are 
mainly responsible for the excess of the quiescent fraction in clusters compared 
to the field region may be gradual, time-taking processes --- like the strangulation 
\citep{lar80,bal00} --- rather than abrupt or violent processes --- like 
ram-pressure stripping, for low-mass galaxies.
The investigation of the morphology of red SF galaxies --- 
which shows that the most transition from the blue SF galaxies to 
red quiescent ones in clusters does not affecting their morphological 
appearance significantly --- supports this idea.

One interesting question is why the SFHs of galaxies are affected 
by their stellar mass. 
More specifically, why is the quiescent galaxy fraction higher for more massive 
galaxies at all environment over a wide range of redshift ($z \lesssim 2$) 
and without strong environmental dependence?
One possible answer is that the dynamical time-scale would be shorter for the 
galaxies with greater stellar masses than less massive galaxies. 
If the SFR of galaxies depends on this dynamical time-scale, in a sense that 
shorter dynamical time-scale leads to higher SFR --- as assumed in many 
galaxy formation models \citep[e.g.,][]{som12}, more massive galaxies will 
consume their gas more rapidly to arrive the red quiescent galaxy sequence 
earlier than low mass galaxies. 
Major mergers between massive SF galaxies can also accelerate this fast gas 
consumption. 
Another possibility is that any negative feedback which works preferentially 
for massive galaxies --- such as AGN feedback \citep[e.g.][]{hop06,som08,san14} or 
halo mass quenching \citep{bir03} --- is the driver of the 
mass-quenching phenomena.

\begin{figure}[h]
\figurenum{A1}
\plotone{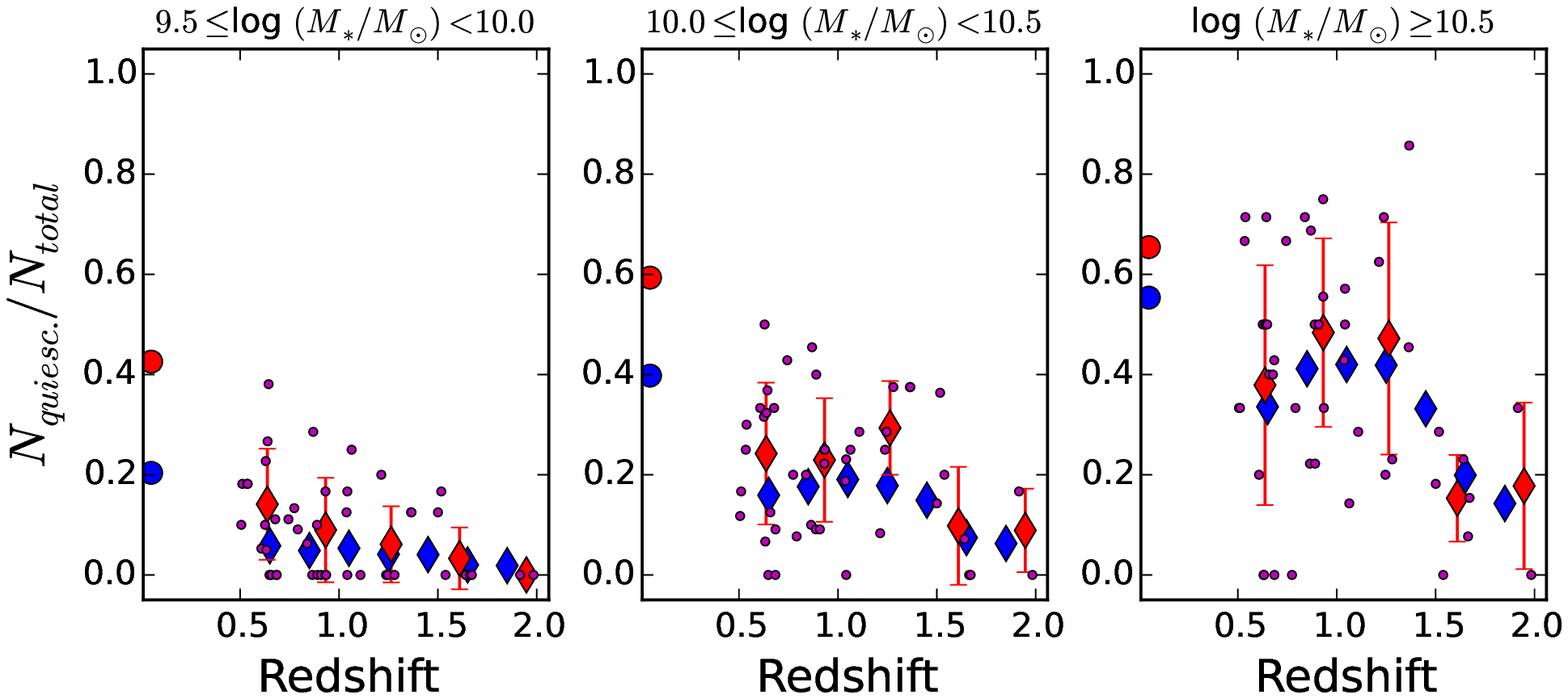}
\caption{The quiescent fraction evolution of galaxies in three 
stellar-mass bins (log $M_{*}/M_{\odot} \sim 9.75$ (left), 
10.25 (middle), and $\geq 10.5$ (right) when we apply constant sSFR 
cut (sSFR = $10^{-10.7}$ yr$^{-1}$). 
The symbol assignment is same as in Figure~\ref{psfracmsbin}.
Similarly with Figure~\ref{psfracmsbin}, the difference of quiescent 
fraction in different stellar mass bins is greater 
than the difference between different environment in the same stellar mass bin 
in both cases. \label{qfevol}}
\end{figure}

One of the interesting results of this work is that the quiescent galaxy 
fraction drops rapidly as we approach the redshift, $z \sim 2$, independent 
of the environment in which they reside. 
This result indicates that the majority of galaxies are actively forming 
stars at $z \gtrsim 2$. 
Combined with the fact that the quiescent fraction within clusters and 
in the field are similar at this epoch, this indicates that studies of clusters 
or proto-clusters at $z \gtrsim 2$ would be crucial in revealing the very 
initial properties of the forming galaxy clusters as well as the evolution 
of galaxies in these massive structures. 
Also, we can expect that any cluster finding methods using the presence of old and 
quiescent galaxies within clusters (for example, like red-sequence technique) would 
miss many clusters or proto-clusters at $z \gtrsim 2$. 
In our work, both of the identification of the galaxy clusters at 
$z \gtrsim 1.5$, as well as the reliable estimation of the SFR of galaxies 
through the SED-fitting are possible thanks to the deep NIR 
data from the UKIDSS. 
This implies that reliable NIR (photometric or spectroscopic) data are 
essential for the future search and the study of the high-redshift galaxy 
clusters during this important epoch ($z  > 1.5$). 
Therefore, we can expect a big leap in our understanding of the properties and 
the evolution of the high-redshift galaxy clusters or proto-clusters as well 
as of the galaxy evolution with the near-future NIR facilities either in space 
--- like JWST (James Webb Space Telescope) --- or on the ground --- like 
GMT (Giant Magellan Telescope) with the GMACS (with the NIR-arm) and later 
with the NIRMOS. 

\begin{figure}[h]
\figurenum{A2}
\plotone{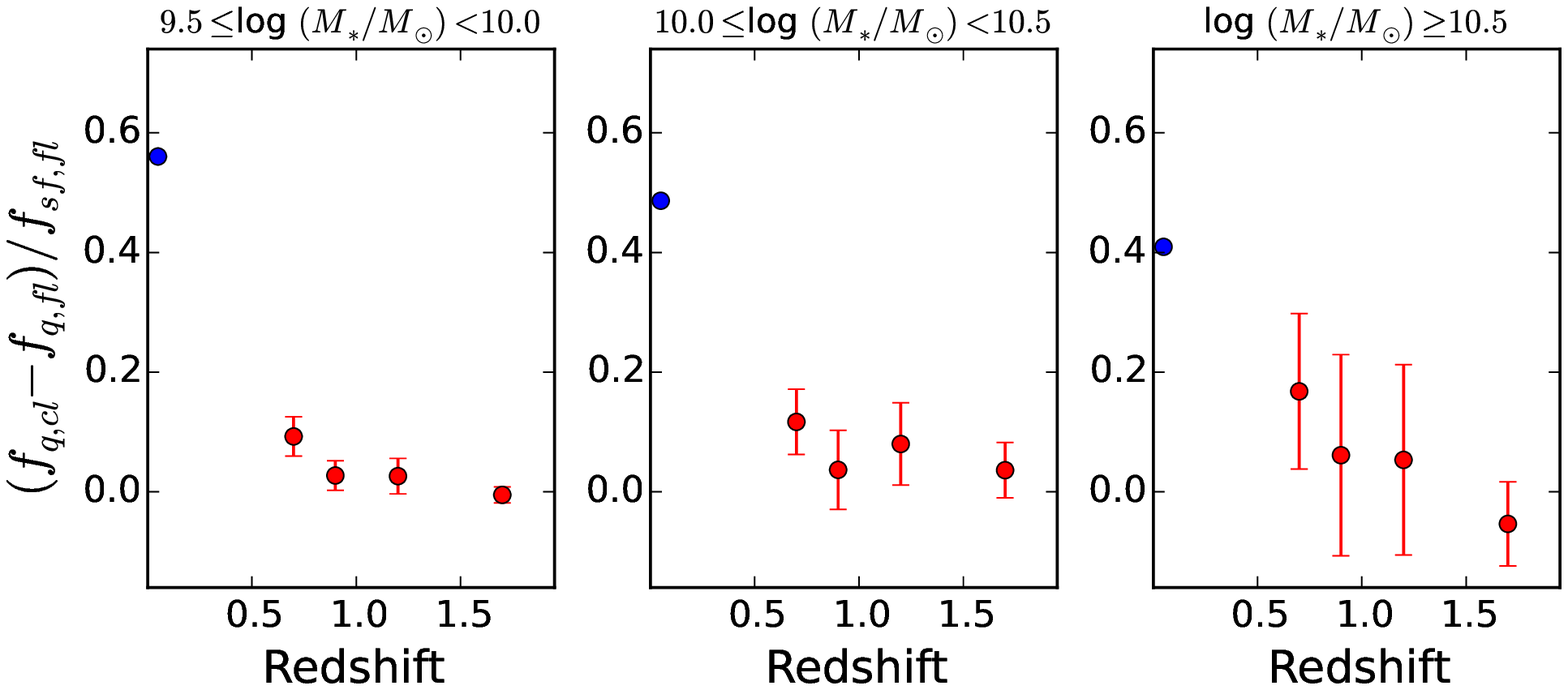}
\caption{Evolution of environmental quenching efficiency, defined as the 
excess of quiescent galaxy fraction in clusters over field divided by SF galaxy 
fraction in field with constant sSFR cut (sSFR = $10^{-10.7}$ yr$^{-1}$). 
The symbol assignment is same as in Figure~\ref{psdiffevol}.
Three panels show this evolution in different stellar mass bins, and 
there is no clear difference between different stellar mass bins. 
\label{qecomp}}
\end{figure}

\acknowledgments

This work was supported by the National Research Foundation of Korea (NRF) grant, 
No. 2008-0060544, funded by the Korea government (MSIP).
The UKIDSS project is defined in \citet{law07}. The UKIDSS uses the UKIRT Wide 
Field Camera \citep[WFCAM;][]{cas07}. The photometric system is described 
in \citet{hew06}, and the calibration is described in  \citet{hod09}. 
The pipeline processing and science archive are described in Irwin 
et al (2009, {\it in prep}) and \citet{ham08}. 
Part of this work is based on observations taken by the CANDELS Multi-Cycle 
Treasury Program with the NASA/ESA HST, which is operated by the Association 
of Universities for Research in Astronomy, Inc., under NASA contract NAS5-26555. 

{\it Facilities:} \facility{UKIRT (WFCAM)}, \facility{Subaru (SUPRIMECAM)}, 
\facility{HST (WFC3)}.

\appendix

\section{Effects of different sSFR cut}

Since galaxies selected with our redshift-dependent cut of $sSFR < 1/3t$ are 
not quite ``quiescent" (i.e., no SF) at high redshift, we examine here how a 
constant sSFR cut changes our main conclusion regarding SF quenching. 
Here, we choose a cut of sSFR = $10^{-10.7}$ yr$^{-1}$, which corresponds to 
the cut for local quiescent galaxies 
\citep[e.g.][]{gal09,ko14}. 

Applying this constant sSFR cut, we first analyse the evolution of 
quiescent galaxies fraction. 
In Figure~\ref{qfevol} --- which corresponds to upper panels of Figure 18, 
we show the evolution of quiescent galaxy fraction in different mass bins. 
Not surprisingly, the quiescent galaxy fraction is overall lower than when 
we apply the 1/[$3 t(z)$]-cut.
The most significant difference is that the quiescent galaxy fraction 
shows steeper increase from $z \sim 2$ to $z \sim 1.3$, for galaxies with 
$M_{*} \geq 10^{10} M_{\odot}$ (middle and right panels), especially 
for $M_{*} \geq 10^{10.5} M_{\odot}$, than in Figure 18. 
This indicates that SF quenching (to local quiescent galaxy level) 
is faster for more massive galaxies.
This fast increase of quiescent fraction at high-$z$ agrees 
with \citet{dom11} result.

While there are interesting differences in quiescent fraction evolution with 
different sSFR cuts, one important trend seems to hold: the quiescent galaxy fraction 
shows clearer dependence on their stellar mass than the environment.
The environmental quenching efficiency (Figure~\ref{qecomp}) is not 
mass-dependent, but rises toward lower redshift (at $z < 1$), similarly with 
Figure 20.

\clearpage

\begin{deluxetable}{ccccccl}
\tablecolumns{7} \tablewidth{0pc} \tablecaption{Candidate high-redshift galaxy clusters in the UDS \label{tab1}}
\tablehead{ \colhead{RA }   &
\colhead{dec. }   &
\colhead{$z$ }   &
\colhead{$N_{gal}$ }   &
\colhead{$\Sigma M_{*}$ }   &
\colhead{$\sigma_{OD}$}   &
\colhead{Reference} \\
\colhead{(1)} & \colhead{(2)} & \colhead{(3)} & \colhead{(4)} & 
\colhead{(5)} & \colhead{(6)} & \colhead{(7)}}
\startdata
 34.48571  &  -4.88290 &  0.506 & 46 &  7.08 &  7.81 & 1 \\
 34.66831  &  -5.05773 &  0.511 & 39 & 10.42 &  7.25 &  \\
 34.05109  &  -4.74287 &  0.534 & 42 &  5.61 &  6.30 & \\
 34.70426  &  -5.14720 &  0.538 & 34 &  6.80 &  5.61 & \\
 34.54121  &  -5.36501 &  0.607 & 55 &  6.44 &  4.81 & 2 \\
 34.54105  &  -5.26098 &  0.626 & 63 & 10.48 &  6.37 & \\
 34.35041  &  -5.41133 &  0.629 & 57 &  6.88 &  5.74 & 1 \\ 
 34.19640  &  -5.15057 &  0.633 & 59 &  7.57 &  4.89 & 1,2 \\
 34.39710  &  -5.22284 &  0.639 & 93 & 16.34 & 13.60 & 1,2 \\ 
 34.60152  &  -5.41210 &  0.644 & 79 & 14.49 &  8.78 & 1 \\ 
 34.51734  &  -5.52098 &  0.648 & 46 &  6.90 &  6.50 & \\
 34.63576  &  -4.96694 &  0.659 & 39 &  5.20 &  5.23 & 2 \\
 34.47932  &  -5.45278 &  0.677 & 60 & 11.63 & 10.05 & 1 \\
 34.74237  &  -5.12341 &  0.684 & 32 &  6.54 &  6.37 & \\
 34.37319  &  -4.68903 &  0.685 & 29 &  3.81 &  6.75 & \\
 34.84291  &  -4.82145 &  0.744 & 28 &  3.31 &  5.49 & \\
 34.45845  &  -5.50699 &  0.774 & 42 &  5.43 &  6.73 & \\
 34.42685  &  -5.09312 &  0.791 & 37 &  7.44 &  5.60 & \\
 34.52339  &  -4.73828 &  0.839 & 44 &  9.04 &  5.73 & 1 \\
 34.82770  &  -5.08506 &  0.865 & 39 &  9.12 &  5.92 & 1 \\
 34.63694  &  -5.01183 &  0.869 & 72 & 13.25 &  9.45 & 1 \\
 34.16359  &  -4.73395 &  0.889 & 46 &  7.78 &  5.12 & \\
 34.84166  &  -4.88236 &  0.890 & 53 &  8.08 &  5.12 & 1 \\
 34.34870  &  -5.20672 &  0.909 & 54 &  9.74 &  7.36 &	 \\   
 34.05085  &  -4.87705 &  0.931 & 54 &  8.41 &  6.16 & \\
 34.06289  &  -4.71558 &  0.932 & 51 &  6.93 &  6.50 & \\
 34.53990  &  -5.01259 &  0.935 & 56 &  7.96 &  5.87 & 2 \\
 34.03834  &  -5.10964 &  1.042 & 43 &  8.61 &  5.18 &	  \\  
 34.29435  &  -4.79312 &  1.042 & 27 &  6.09 &  5.09 & 2 \\
 34.59195  &  -4.97802 &  1.038 & 38 &  8.43 &  5.93 & \\
 34.52635  &  -5.01766 &  1.064 & 38 & 10.51 &  5.68 & 2 \\
 34.28226  &  -4.82255 &  1.109 & 34 &  9.19 &  5.68 & 1 \\
 34.61517  &  -4.69712 &  1.214 & 35 &  8.57 &  4.89 & \\
 34.69049  &  -4.71229 &  1.238 & 46 & 10.41 &  5.34 & \\
 34.80897  &  -4.93465 &  1.246 & 43 &  6.03 &  5.73 & \\
 34.53326  &  -5.01095 &  1.281 & 41 & 12.55 &  6.12 & 3 \\ 
 34.85731  &  -4.86560 &  1.364 & 49 & 15.61 &  7.69 &   \\
 34.84134  &  -4.72292 &  1.366 & 32 &  8.73 &  5.63 & \\
 34.53797  &  -5.01744 &  1.500 & 36 &  9.48 &  5.60 & 1,3  \\
 34.07070  &  -5.00289 &  1.517 & 36 &  8.18 &  5.49 & \\
 34.81516  &  -4.74594 &  1.538 & 41 &  7.15 &  6.18 & \\
 34.59165  &  -5.16940 &  1.640 & 66 & 14.37 &  6.93 & 1,4 \\
 34.17832  &  -5.15493 &  1.664 & 62 & 11.41 &  5.34 & \\
 34.80031  &  -4.73156 &  1.671 & 77 & 13.34 &  6.17 & \\
 34.73531  &  -5.04661 &  1.916 & 40 &  4.73 &  6.89 & \\
 34.72378  &  -5.17741 &  1.983 & 35 &  6.79 &  5.30 & \\
\enddata

\tablecomments{\\ (1) RA in degree \\ (2) Declination in degree \\ (3) Redshift \\
(4) Number of galaxies within 1 Mpc radius from the cluster center \\  
(5) Sum of the stellar masses of the member galaxies in 
1e+11 $M_{\odot}$ \\ (6) Overdensity measure, ($N - \bar{N}$)/$\sigma_{N}$ \\  
(7) Reference list: \\ ~~~~~~1. Finoguenov et al. 2010, $MNRAS$, 403, 2063 \\ 
~~~~~~2. van Breukelen et al. 2006, $MNRAS$, 373, L26 \\ ~~~~~~3. van Breukelen et al. 2007, $MNRAS$, 382, 971\\ ~~~~~~4. Papovich et al. 2010, $ApJ$, 716, 1503 }


\end{deluxetable}

\clearpage

\end{document}